\documentclass[12pt]{iopart}
\usepackage{iopams}  
\usepackage{graphicx}
\usepackage[dvipsnames]{xcolor}

\newcommand{\rhotor}{\rho_{\rm tor}}
\newcommand{\fig}[1]{Fig.~\ref{#1}}
\newcommand{\kxcenter}{k_{\rm x,center}}
\newcommand{\Apar}{\bar{A}_{1\parallel}}
\newcommand{\Bpar}{\bar{B}_{1\parallel}}

\begin{document}

\title[The JET hybrid H-mode scenario from a pedestal turbulence perspective]{The JET hybrid H-mode scenario from a pedestal turbulence perspective}

\author{L. A. Leppin$^1$, T. Görler$^1$, L. Frassinetti$^2$, S. Saarelma$^3$, J. Hobirk$^1$, F. Jenko$^1$ and JET contributors$^4$}

\address{
$^1$ Max Planck Institute for Plasma Physics, Boltzmannstraße 2, 85748 Garching b. München, Germany\\
$^2$ Division of Fusion Plasma Physics, KTH Royal Institute of Technology, SE-10691 Stockholm, Sweden \\
$^3$ CCFE, Culham Science Center, Abingdon OX14 3DB, United Kingdom of Great Britain and Northern Ireland \\
$^4$ See C. Maggi et al 2024 (https://doi.org/10.1088/1741-4326/ad3e16) for the JET Contributors
}
\ead{leonhard.leppin@ipp.mpg.de}
\vspace{10pt}
\begin{indented}
\item[]\today
\end{indented}

\begin{abstract}
Turbulent transport is a decisive factor in determining the pedestal structure of H-modes. Here, we present the first comprehensive characterization of gyrokinetic turbulent transport in a JET hybrid H-mode pedestal. Local, linear simulations are performed to identify instabilities and global, nonlinear electromagnetic simulations reveal the turbulent heat and particle flux structure of the pedestal. Our analysis focuses on the Deuterium reference discharge \#97781 performed in the scenario development for the Deuterium-Tritium campaign. We find the pedestal top transport to be dominated by ion temperature gradient (ITG) modes. In the pedestal center turbulent ion transport is suppressed and electron transport is driven by multi-faceted electron temperature gradient (ETG) modes, which extend down to ion-gyroradius scales. A strong impact of $E\times B$ shear on the absolute turbulence level is confirmed by the global, nonlinear simulations. Furthermore, impurities are shown to reduce the main ion transport. Dedicated density and ion temperature profile variations test the sensitivity of the results and do not find strong differences in the turbulent transport in more reactor-like conditions.
\end{abstract}

%
\vspace{2pc}
\noindent{\it Keywords}: JET, tokamak, pedestal, turbulence, gyrokinetics, hybrid, H-mode

%
\submitto{\NF}
%
\maketitle
%
%

\section{Introduction: JET hybrid H-mode scenario}
The high-confinement mode (H-mode) is an attractive operating scenario for tokamak fusion devices \cite{wagner_regime_1982,wagner_quarter-century_2007}. It is the default high-performance scenario of ITER \cite{iter_physics_expert_group_on_confinement_and_transport_chapter_1999} and other next-step fusion devices like SPARC \cite{creely_overview_2020}. H-mode discharges are characterized by an edge transport barrier, which causes large temperature and density gradients in the outer few centimeters of the confined plasma, the so-called pedestal. 

At the JET tokamak, the H-mode scenario has been optimized in a particular kind of discharge, namely hybrid discharges \cite{joffrin_hybrid_2005,hobirk_improved_2012,beurskens_comparison_2013,challis_improved_2015,hobirk_jet_2023}. Hybrids are an operating scenario located in parameter space between the pulsed baseline scenario (high inductive plasma current, low normalized plasma $\beta_N=\beta_TB_T/aI_p$ (in JET $\beta_{N \rm{,baseline}}\approx1.5$)) and advanced tokamak scenario (non-inductive plasma current, high $\beta_N$) aiming at steady-state operation. Typical values in JET hybrids are $\beta_{N\rm{,hybrid}}\approx2.5$. Hybrid H-modes feature a higher H-factor ($H_{98,y2}\approx1.2 - 1.4$) compared to the baseline scenario $H_{98,y2}\approx1$. Recent JET performance records were achieved in hybrid H-modes \cite{hobirk_jet_2023,maslov_jet_2023}, making hybrid H-modes an attractive operating scenario for ITER.

These experimental findings raise two important questions: Why is the confinement in the hybrid scenario improved compared to the baseline scenario? And can the current hybrid scenarios be extrapolated to larger machines like ITER? The complete answers to these questions most likely involve many interacting aspects of the plasma discharge, from core to edge physics and from the smallest turbulent scales to large-scale MHD effects. Several contributing mechanisms have been identified in previous work: Fast ions have been found to reduce core turbulence \cite{garcia_key_2015,moradi_core_2014,baiocchi_turbulent_2015} and increase pedestal pressure through MHD effects \cite{garcia_key_2015}. Furthermore, an electromagnetic and/or shear stabilization of ion temperature gradient (ITG) turbulence in the core has been reported \cite{citrin_electromagnetic_2015,doerk_gyrokinetic_2016,moradi_core_2014,mariani_benchmark_2021}. So far studies have focused on gyrokinetic turbulence in the core of hybrids ($\rhotor<0.7$). Experimental analysis, however, suggests that both core and edge contribute to the improved confinement of hybrids \cite{hobirk_improved_2012,challis_improved_2015}. This study aims to close this gap, by a detailed gyrokinetic analysis of a hybrid H-mode pedestal. From a gyrokinetic pedestal perspective the present study complements previous studies on JET baseline H-mode pedestals, which focus on the difference in transport between carbon wall and ITER-like wall \cite{hatch_direct_2019,hatch_gyrokinetic_2017}, electron temperature gradient (ETG)  turbulence \cite{parisi_toroidal_2020,chapman-oplopoiou_role_2022,field_comparing_2023} and isotope effects \cite{predebon_isotope_2023}.

Here, we present the first comprehensive investigation of pedestal turbulence ($\rhotor=0.85-0.995$) in a JET hybrid H-mode. We investigate small-scale instabilities and turbulence structure with local/linear, local/nonlinear, and global/nonlinear gyrokinetic simulations. Global/nonlinear simulations employ a recent upgrade of the GENE code \cite{leppin_complex_2023-1} enabling global, nonlinear simulations at experimental plasma $\beta$ values. Starting from a turbulence characterization at nominal best-fit parameters we assess the influence of density and temperature variations on the turbulent fluxes. These profile variations probe the sensitivity of our results and address the discussed questions on the reasons for the increased H-factor of hybrid discharges and its viability as an ITER scenario. We find turbulent transport to be mostly driven by ion temperature gradient (ITG) modes at the pedestal top and electron temperature gradient (ETG) modes with a multi-faceted character in the pedestal center. Both ion- and electron-scale transport show quasi-linear properties. The turbulent state is found to be robust against changes in collisionality, but sensitive to changes in the plasma $\beta$. Changes in collisionality are not found to explain the improved pedestal confinement properties of hybrid H-modes. Profile variations to more reactor-like conditions with a temperature ratio $T_i/T_e=1$ or further reduced collisionality do not cause a strong change or increase in the turbulent heat flux level.

This paper is structured as follows: Section \ref{sec:exp_scenario} introduces the experimental shot JET \#97781 investigated in this paper. The simulation setups are detailed in Section \ref{sec:simsetup}. Section \ref{sec:nominal} then presents a characterization of instabilities and turbulence at nominal parameters including local/linear growth rate scans, local/nonlinear ETG simulations, and global/nonlinear ion scale simulations. Heat and particle fluxes are discussed, as well as the quasi-linear nature of pedestal turbulence in this scenario. Section \ref{sec:variation} discusses the influence of selected density and temperature profile changes. Finally, in Section \ref{sec:conclusions} conclusions are drawn.

\section{Experimental scenario: JET \#97781}
\label{sec:exp_scenario}
The basis of our gyrokinetic study is a pre-ELM pedestal of the hybrid H-mode shot JET \#97781 which served as the Deuterium reference discharge in JET's DT scenario development \cite{hobirk_jet_2023}. It features a normalized pressure of $\beta_N=2.34-2.77$ and an H-factor of $H_{98,y2}=1.18 - 1.4$ \cite{hobirk_jet_2023}. In the investigated time window of 8.8 s - 9.9 s this shot has a plasma current of $I_p=2.3$ MA, an on-axis toroidal magnetic field of $B_t=3.45$ T, heating by neutral beams (NBI) of $P_{\rm NBI}=30$ MW and ion cyclotron heating (ICRH) of $P_{\rm{ICRH}}=3$ MW.

\begin{figure}
    \centering
    \includegraphics[width=0.9\linewidth]{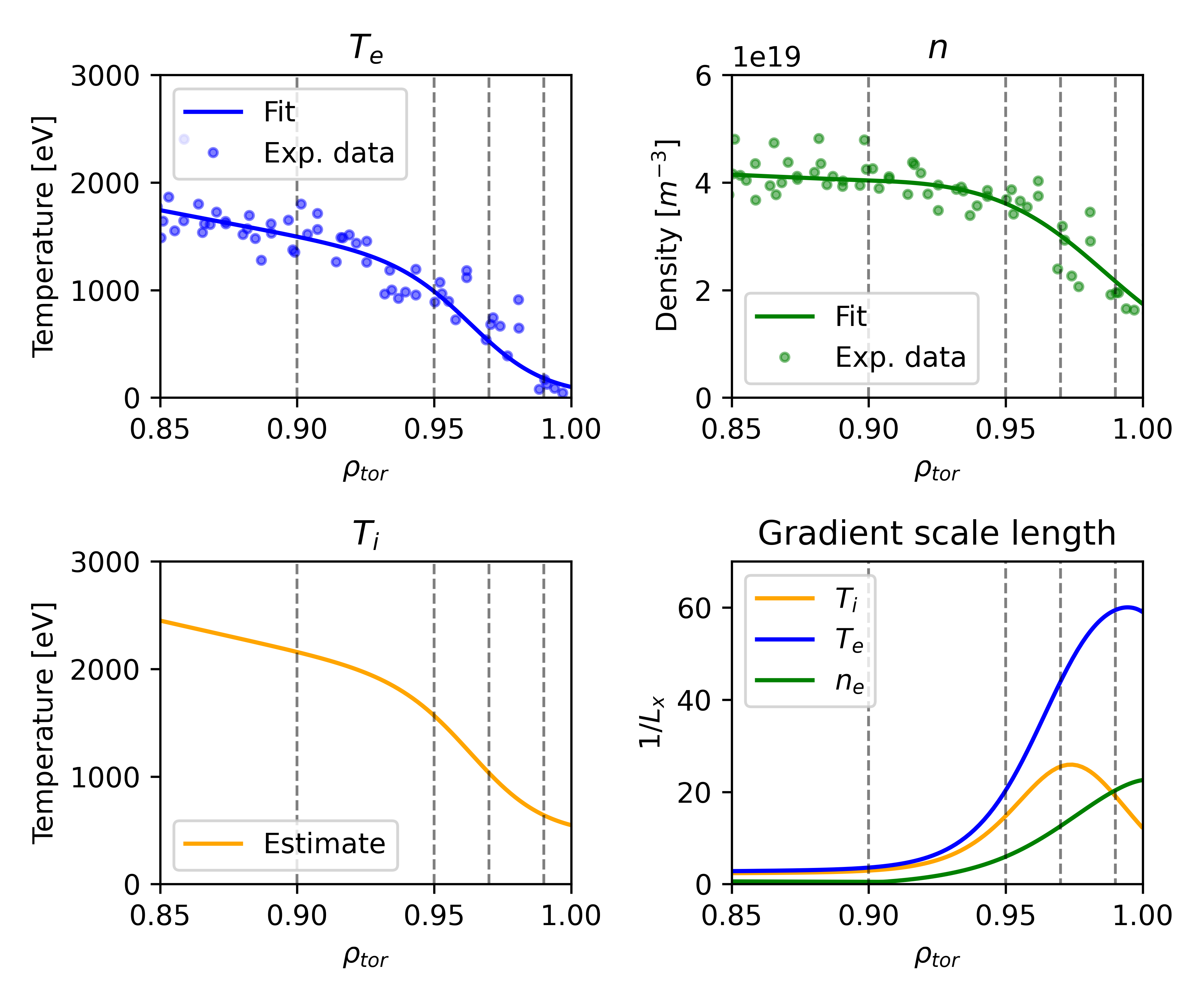}
    \caption{Temperature and density pedestal profiles of JET \#97781 from the time window 8.8 s - 9.9 s. Vertical dashed lines indicate locations analyzed in growth rate scans.}
    \label{fig:profiles}
\end{figure}

Fig.~\ref{fig:profiles} shows the measured pre-ELM electron temperature and density pedestal profiles and the corresponding best fits with a modified hyperbolic tangent \cite{groebner_progress_2001} and determined as described in \cite{frassinetti_spatial_2012}. The radial coordinate $\rhotor$ is the square-root of the normalized toroidal magnetic flux. For the ion temperature profile, it was assumed that $T_{i,ped}=1.5T_{e,ped}$ and $T_{i,sep}=500$ eV. The effect of different assumed ion temperatures ($T_i=T_e$ and $T_i=1.3T_e$) is discussed in Section \ref{sec:temp_variation}. The ion density is assumed to be equal to the electron density, except if an impurity species is explicitly assumed (see discussion in Section \ref{sec:nominal}). In this case, the main ion density is assumed to be reduced such that quasi-neutrality is fulfilled. The gradient scale lengths corresponding to the nominal profiles are shown in the lower right panel of Fig.~\ref{fig:profiles}.

\begin{figure}
    \centering
    \includegraphics[width=0.54\linewidth]{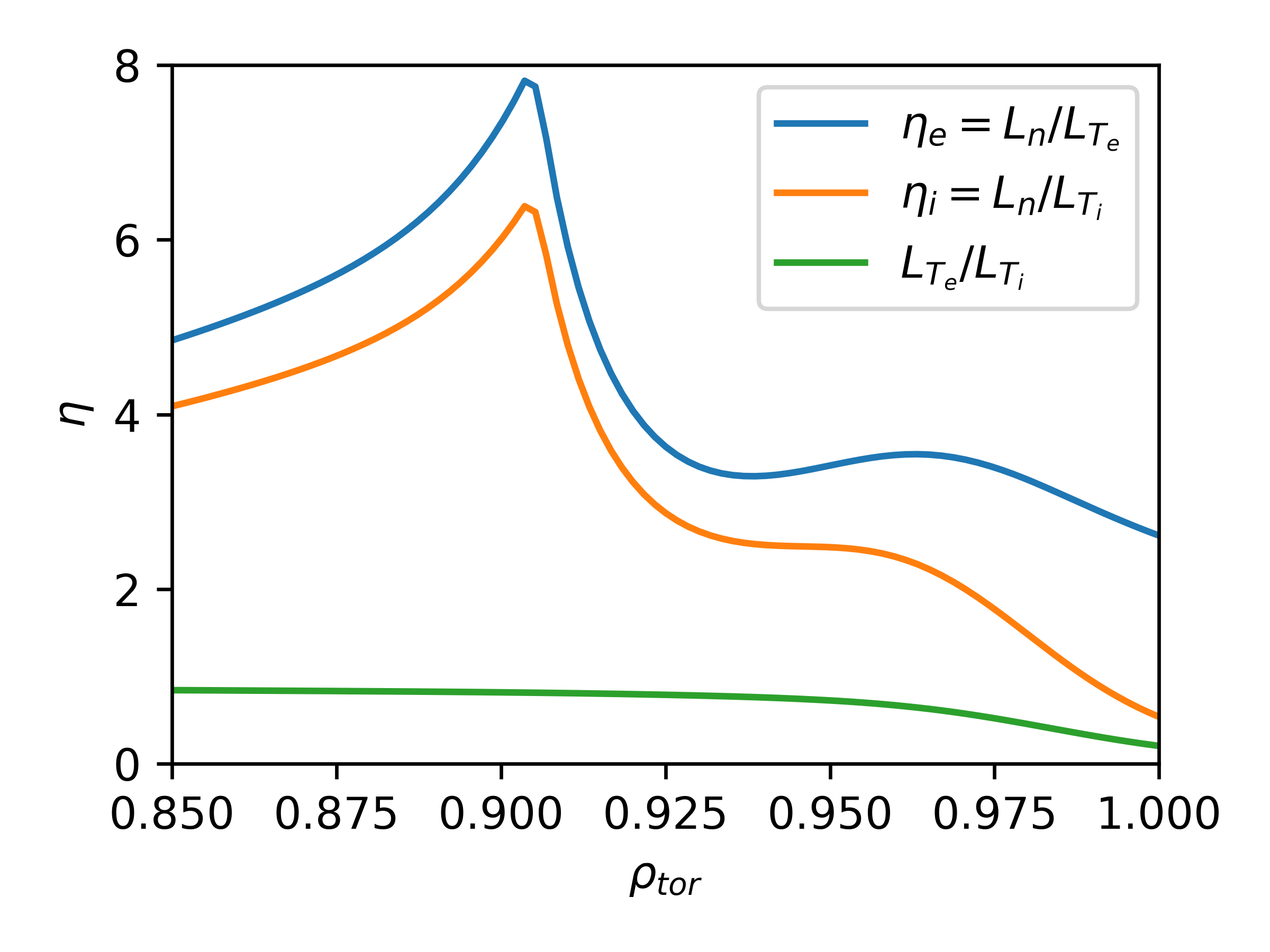}
    \caption{Gradient ratios and temperature ratio in the pedestal of JET \#97781.}
    \label{fig:derived_profiles}
\end{figure}

A first insight into potential gyrokinetic instabilities can be gained by considering the ratios of density and temperature gradients $\eta_e=L_n/L_{T_e}$ and $\eta_i=L_n/L_{T_i}$. Fig.~\ref{fig:derived_profiles} shows these ratios as a function of radius. Their profiles indicate a strong drive of ion temperature and electron temperature gradient (ITG, ETG) instabilities throughout the pedestal, particularly around $\rhotor=0.9$, due to a small density gradient compared to the temperature gradients.

\begin{figure}
    \centering
    \includegraphics[width=0.9\linewidth]{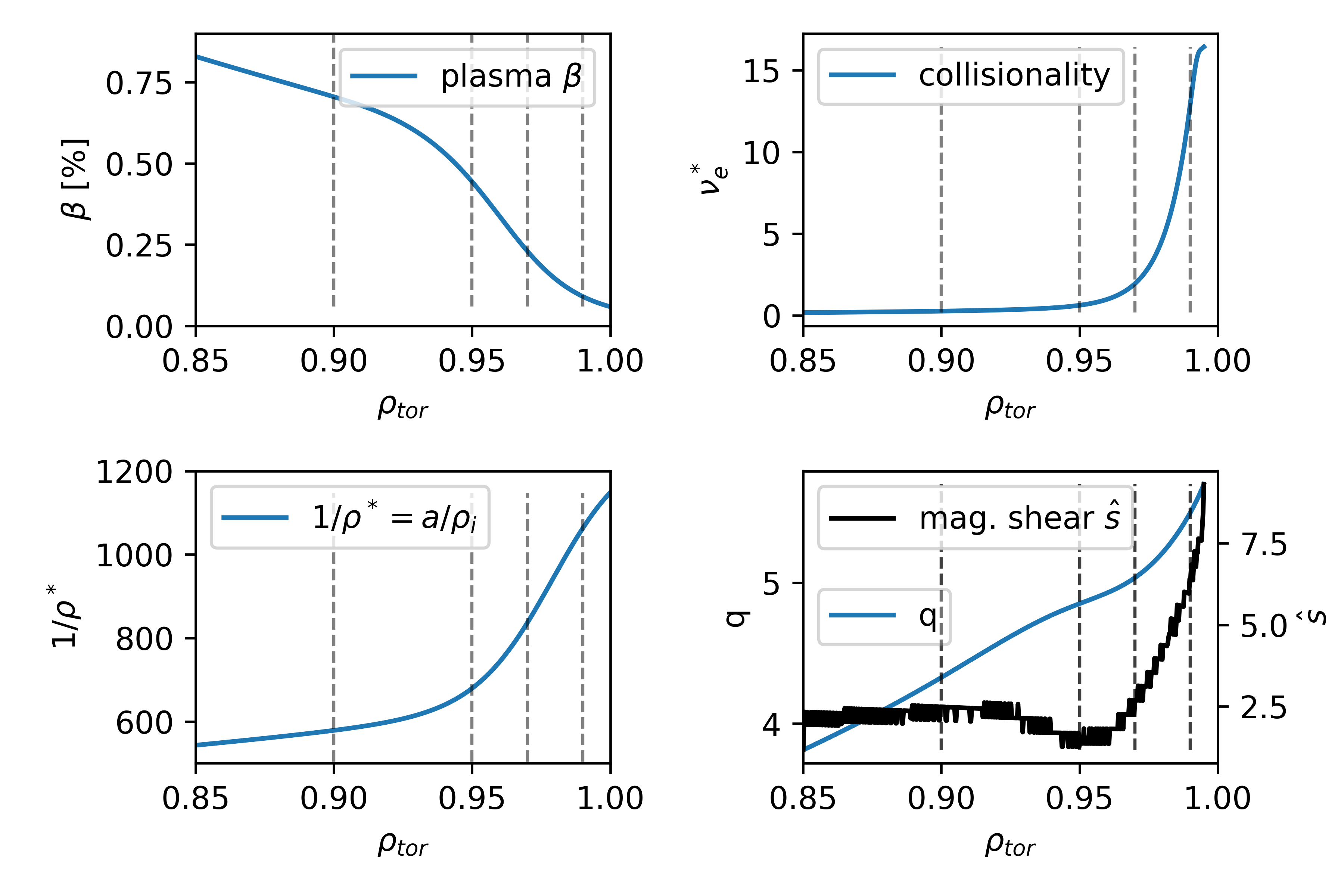}
    \caption{Radial profiles of further important pedestal quantities.}
    \label{fig:more_profiles}
\end{figure}

Further important physical quantities influencing instabilities and turbulence are the plasma $\beta$, collisionality, ion gyroradius size, and safety factor/magnetic shear. Their radial profiles are shown in Fig.~\ref{fig:more_profiles}. As in a preceding study on the structure of turbulence in an ASDEX Upgrade pedestal \cite{leppin_complex_2023-1}, the magnetic shear has a minimum (although in the current case less pronounced) at the onset of the steep gradient region at $\rhotor=0.95$ (see lower right panel). An experimental magnetic equilibrium consistent with the kinetic profiles is used.

\begin{figure}
    \centering
    \includegraphics[width=0.9\linewidth]{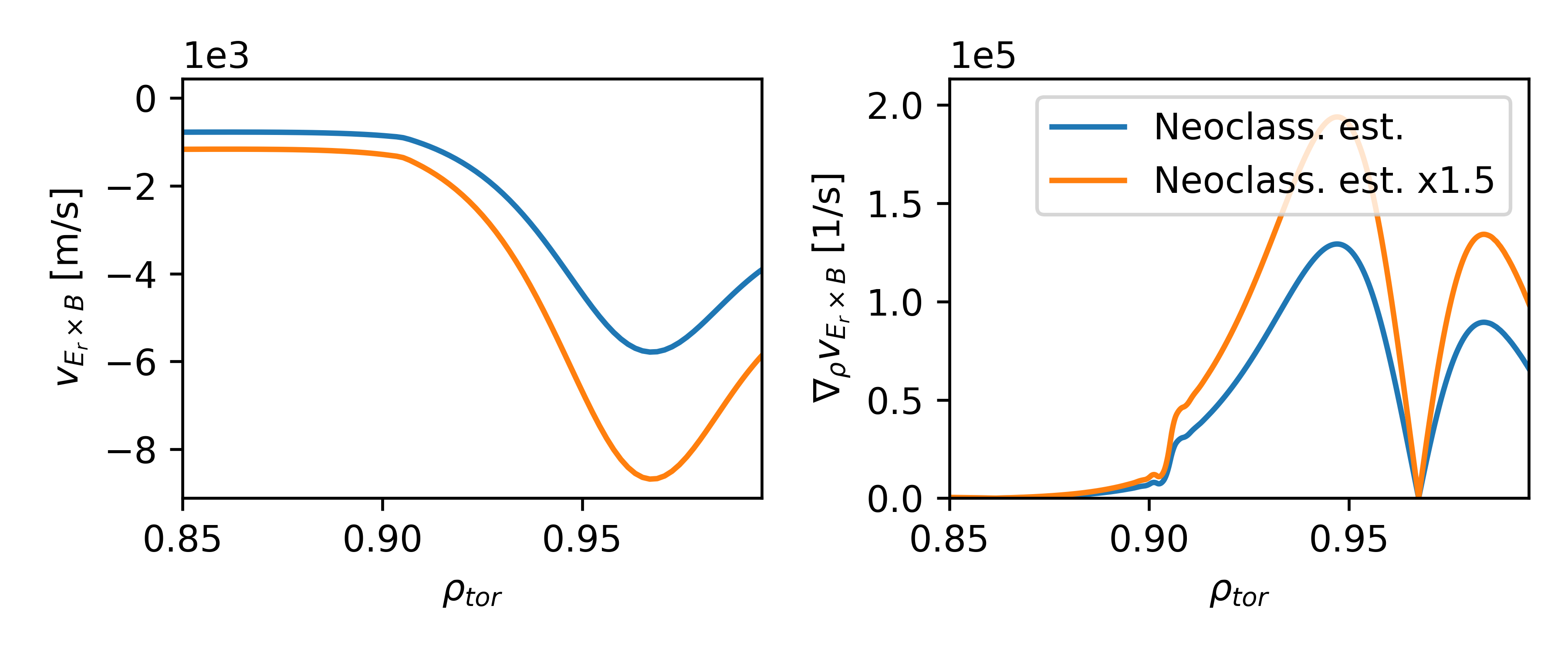}
    \caption{Radial profile of estimated rotation velocity from radial force balance (left) and corresponding shear (right).}
    \label{fig:vrot}
\end{figure}

$E\times B$ shear is known to suppress ion-scale turbulence and is thought to be an important factor in pedestal dynamics. In this study, we use (neoclassic) radial force balance to estimate the radial electric field and corresponding rotation velocity, following Ref.~\cite{hatch_gyrokinetic_2017}. The calculation is based on a script developed and employed in earlier JET studies with GENE \cite{hatch_gyrokinetic_2017,hatch_direct_2019,chapman-oplopoiou_role_2022,predebon_isotope_2023}. Fig.~\ref{fig:vrot} shows the obtained rotation velocity (left) and associated shear (right). An increased rotation velocity tested in nonlinear simulations is shown as well (orange).

The impurity content in the pedestal is challenging to determine and prone to measurement uncertainties. Based on dedicated analysis \cite{sertoli_plasma_2020,sertoli_measuring_2019} we model impurity effects in this study by assuming a fully-ionized Beryllium species causing an effective ion charge of $Z_{\rm eff}=1.6$. Beryllium is selected as the impurity species as it causes the strongest main ion dilution for a given $Z_{\rm eff}$ since it has the lowest atomic number among the impurity candidates Nickel, Tungsten, and Beryllium.

The next section describes the simulation setups applied to this experimental scenario before results at nominal parameters are discussed in Section \ref{sec:nominal}.

\section{Simulation setups}
\label{sec:simsetup}
All simulations were performed with the gyrokinetic GENE code \cite{jenko_electron_2000,gorler_global_2011}, some using a recent upgrade of the global, electromagnetic code version \cite{leppin_complex_2023-1}. They are gradient-driven, include perpendicular electromagnetic fluctuations ($\Apar$, but not $\Bpar$), collisions (linearized Landau-Boltzmann collision operator), and two species (except when explicitly stated otherwise) with the correct Deuterium-to-electron mass ratio. When indicated they include $E\times B$ shear. They use experimental temperature and density profiles and a corresponding experimental magnetic equilibrium (g-eqdsk obtained with CHEASE) as inputs.

\subsection{Local/linear}
Scans with local/linear simulations were performed at $\rhotor=0.9, 0.95, 0.97, 0.99$, representing the pedestal top, local minimum of the magnetic shear, pedestal center (steepest gradient) and pedestal foot, respectively. They cover ion to electron scales in the binormal wavenumber $k_y\rho_i=0.05 - 350$ and include the radial wavenumber at the outboard midplane $\kxcenter\rho_i=[-40,40]$, which is related to the ballooning angle $\theta_0=\kxcenter/(\hat{s}k_y)$. The simulations are performed with a resolution of $n_x=18$ in the radial direction, $n_z=36$ in the parallel direction and $n_v=32, n_w=16$ in velocity space with size $l_v=3 (2T_0/m)^{1/2}$, $l_w=9T_0/B_{ref}$, where $T_0$ is the temperature of the respective species on the flux surface. Scans in the parallel resolution reveal that higher parallel resolutions of $n_z\approx360$ are necessary to achieve fully converged growth rates at high binormal wavenumbers $k_y\rho_i>40$ (cf.~Fig.~\ref{fig:nz0_scan}). Local simulations use periodic radial boundary conditions. They are performed in double-precision floating-point format.

\subsection{Local/nonlinear}
Local/nonlinear simulations were performed at $\rhotor=0.97$ in the steep gradient region on ETG scales. These simulations use adiabatic ions and kinetic electrons. They include $n_{ky}=64$ binormal wave numbers with $k_{y,min}=3\rho_i$ and $n_x=256$ radial grid points in a box of $l_x=3.65\rho_i$. The velocity space covers $l_v=3 (2T_0/m)^{1/2}$, $l_w=9T_0/B_{ref}$ with $n_v=32, n_w=16$ grid points. The parallel resolution $n_z$ and hyperdiffusion \cite{pueschel_role_2010} have been scanned from $n_z=288$ up to $n_z=2304$, as discussed in Section \ref{sec:nonlinear}. Local simulations use periodic radial boundary conditions. They are performed in single-precision floating-point format.

\subsection{Global/nonlinear}
Global/nonlinear simulations were performed in the radial domain $\rhotor=0.85-0.995$. Most of them are run with two species, but the impact of impurities is assessed in a simulation with three kinetic species. They use $n_x=512$ radial grid points (a convergence test with $n_x=1024$ was performed), $n_{ky}=32$ wavenumbers, and $n_z=48$ parallel grid points. The velocity space covers at the pedestal top $l_v=3.4 (2T_0/m)^{1/2}$, $l_w=13.7 T_0/B_{ref}$, where $T_0$ is the temperature of the respective species in the center of the simulation domain, i.e. the reference flux surface, with $n_v=32, n_w=16$ grid points. By using block-structured grids \cite{jarema_block-structured_2017} the velocity space box size is adapted in the different pedestal regions (e.g. at $\rhotor=0.97$ it is reduced to $l_v=1.9 (2T_0/m)^{1/2}$, $l_w=4.3 T_0/B_{ref}$) for an optimized resolution. Global simulations use Dirichlet boundary conditions at the radial boundaries with buffer zones (5\% percent of the domain at both boundaries), in which the distribution function is damped by fourth-order Krook-type operators \cite{gorler_global_2011}. Global/nonlinear simulations are performed in single-precision floating-point format. In total all global/nonlinear simulations combined required about 140,000 node hours, mostly performed on the Marconi supercomputer at CINECA.

The results obtained with these setups are discussed in the following sections.

\section{Gyrokinetic characterization at nominal parameters}
\label{sec:nominal}
\subsection{Gyrokinetic instabilities}
\label{sec:linear}
\begin{figure}
    \centering
    Full spectra
    \includegraphics[width=\linewidth]{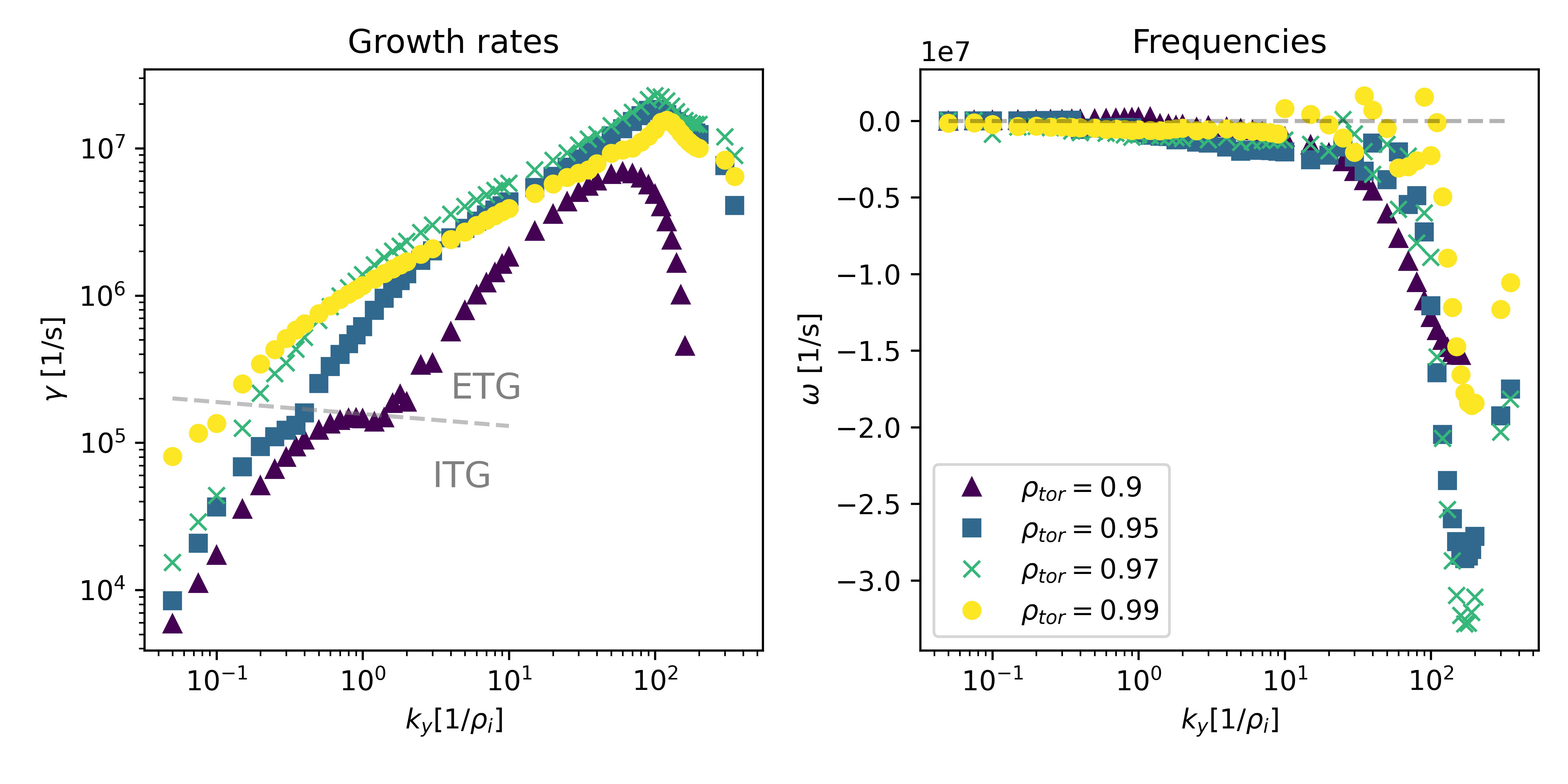}
    Zoom to ion-scales
    \includegraphics[width=\linewidth]{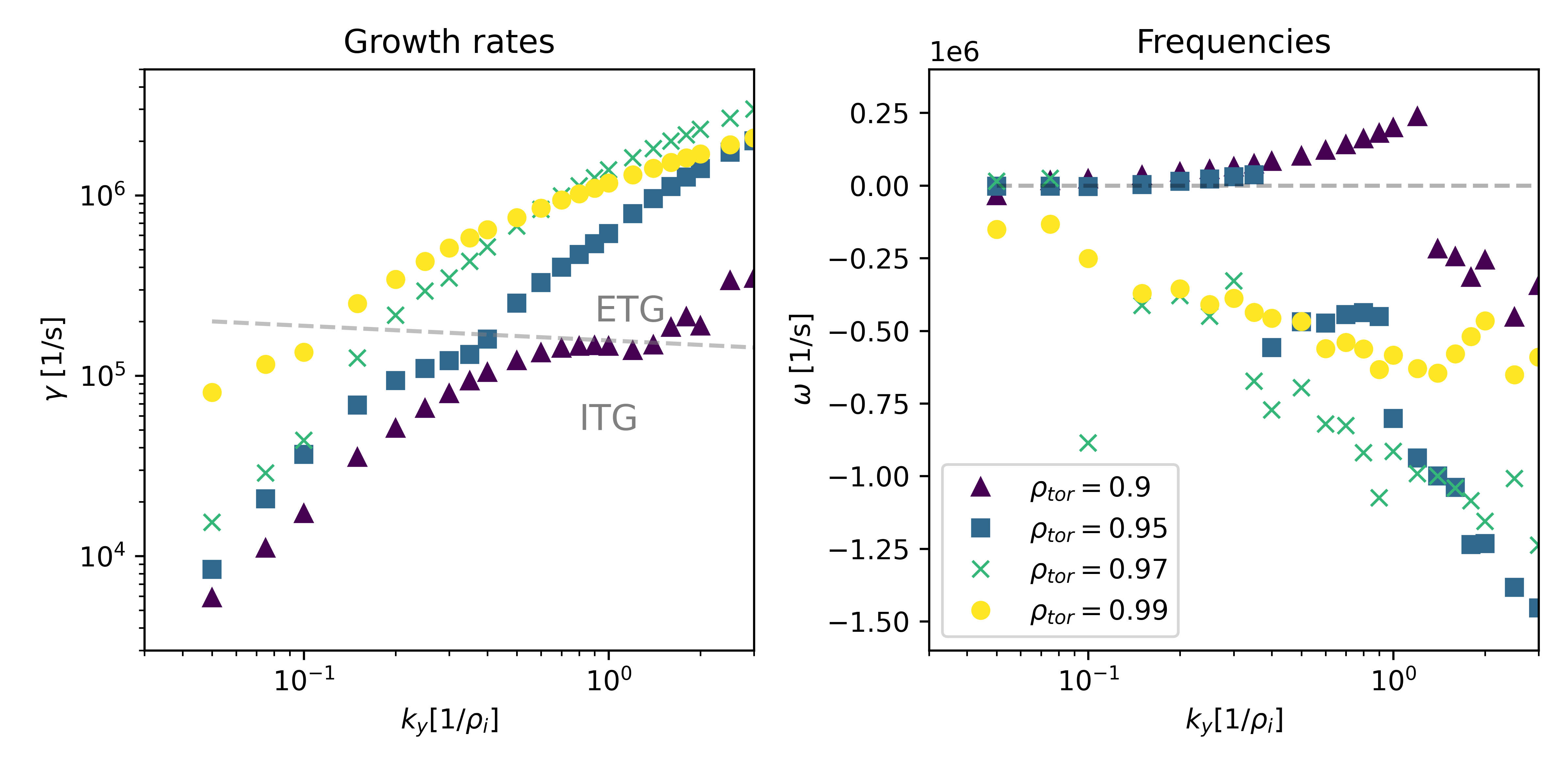}
    \caption{Growth rate and frequency spectra at different radial positions in SI units. Full spectra (top) and zoomed in to ion-gyroradius scales (bottom).}
    \label{fig:gamma_nominal}
\end{figure}

Growth rate and frequency spectra at four radial positions are shown in Fig.~\ref{fig:gamma_nominal}. The pedestal top ($\rhotor=0.9$, purple triangles) on ion scales is dominated by an electrostatic mode propagating in ion-diamagnetic direction (in GENE's sign convention positive/negative frequencies correspond to the ion-/electron-diamagnetic direction, respectively) and ballooning parity that is destabilized when increasing the ion temperature gradient and causes more ion than electron transport. It is identified to be an ITG mode. On smaller scales ($k_y\rho_i>1$) an ETG mode is dominant. It is noteworthy that the spectrum even at the pedestal top does not exhibit a region of stable wavenumbers - eddies of all sizes from ion to electron scales are excited. At $\rhotor=0.95$ (blue squares) the transition of the dominant mode from ITG to ETG is shifted to a lower binormal wavenumber of $k_y\rho_i\approx0.4$.  This trend continues to the pedestal center and foot, where the ETG mode becomes the dominant instability starting from $k_y\rho_i\approx0.15$. Only the largest scale modes $k_y\rho_i<0.15$ remain ITG dominated. None of the modes observed in these spectra of the fastest-growing modes have an electromagnetic character or tearing parity, excluding Kinetic Ballooning Modes (KBMs) and Micro Tearing Modes (MTMs). These modes may, however, still be present sub-dominantly. The inclusion of $\Bpar$ fluctuations in the local/linear simulations was tested, but only very small changes in growth rates were observed compared to the standard GENE $\nabla B$ drift treatment mimicking the MHD-type cancellation of parallel magnetic fluctuations with pressure gradient contributions as discussed, e.g., in Ref.~\cite{waltz_ion_1999}.

The bottom panels highlight the ion-scale instabilities that are responsible for substantial parts of heat transport, particularly at the pedestal top. Specifically, the zoomed-in frequency plot (bottom right panel) allows one to identify mode transitions. The pedestal top $\rhotor=0.9$ (violet triangles) is ITG dominated (positive frequency branch) up to $k_y\rho_i\approx1$ before an ETG branch takes over. At $\rhotor=0.95$ (blue squares), the transition happens earlier ($k_y\rho_i\approx0.4$), and at the pedestal center $\rhotor=0.97$ (green crosses) even earlier ($k_y\rho_i\approx0.08$). At the pedestal foot (yellow circles), no positive ITG frequencies are visible.

\subsubsection{ETG character}
\begin{figure}
    \centering
    \includegraphics[width=0.5\linewidth]{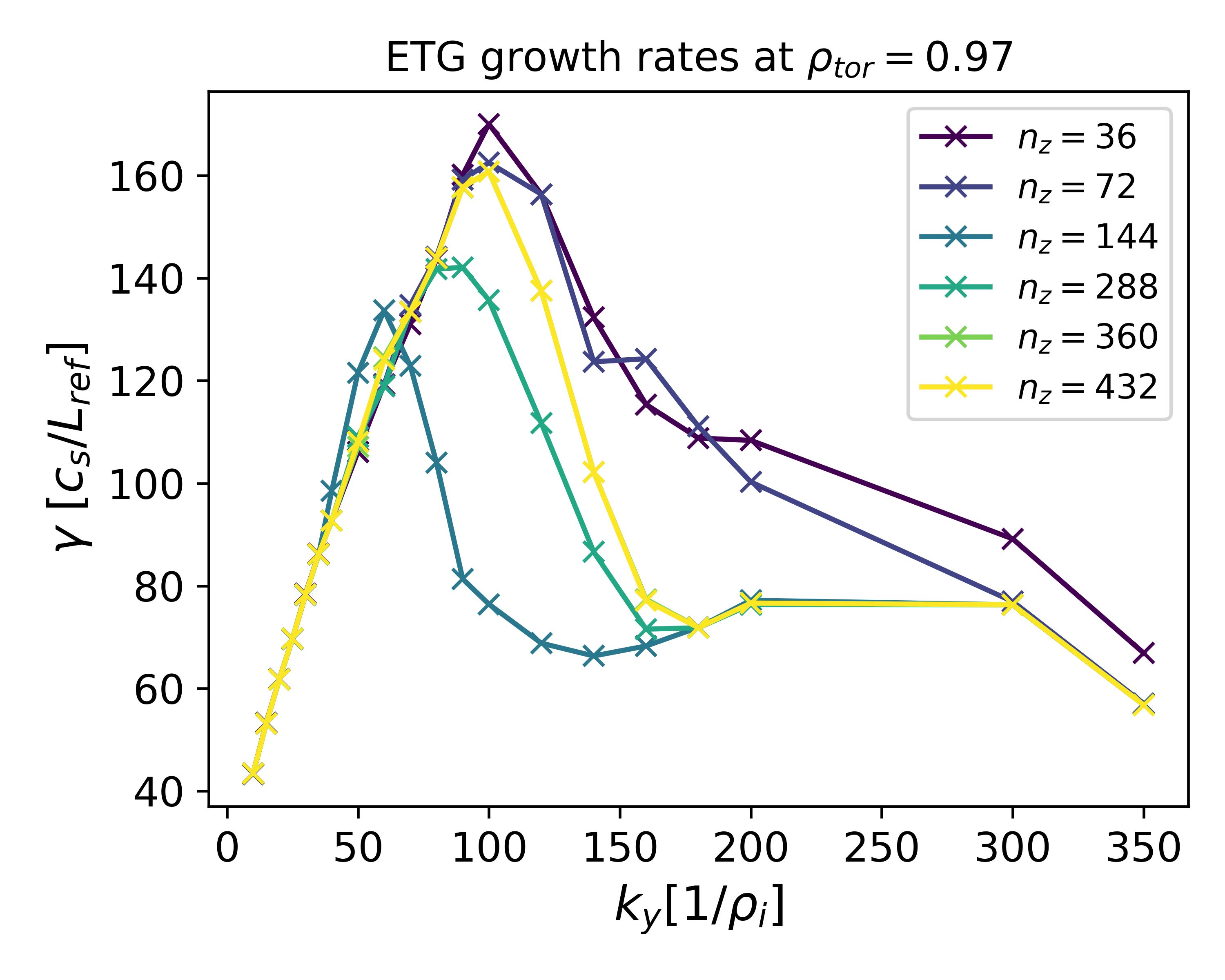}
    \caption{ETG growth rates at $\rhotor=0.97$ for different parallel resolutions $n_z$.}
    \label{fig:nz0_scan}
\end{figure}

\begin{figure}
    \centering
    \includegraphics[width=\linewidth]{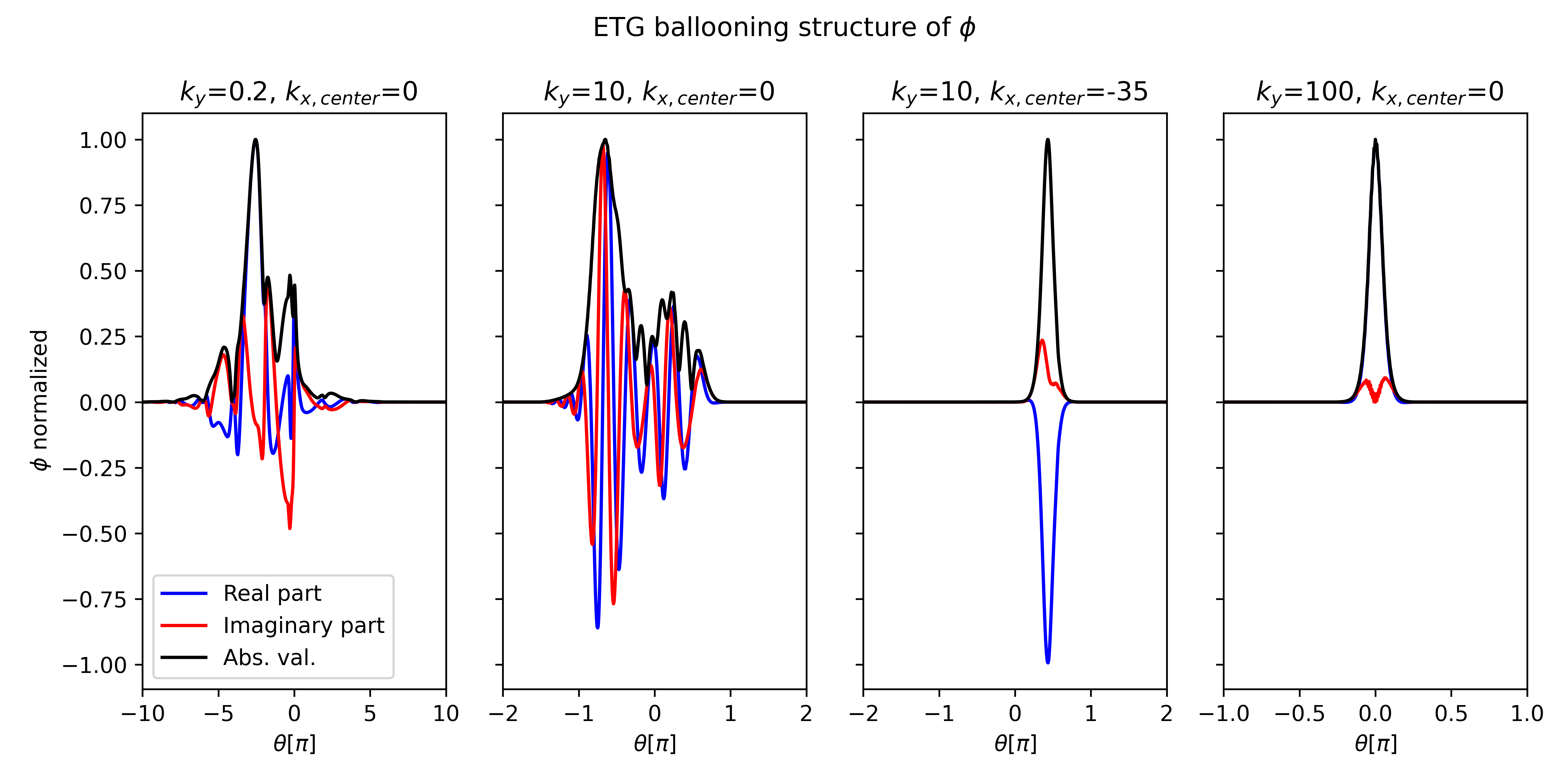}
    \caption{ETG ballooning structures at different binormal wavenumbers $k_y$ and radial wavenumbers $\kxcenter$ in the pedestal center $\rhotor=0.97$.}
    \label{fig:ballooning}
\end{figure}

Since ETG modes are thought to be responsible for significant transport in the pedestal \cite{jenko_electron_2000,kotschenreuther_gyrokinetic_2019,guttenfelder_testing_2021}, it is worthwhile to examine the broad spectrum of ETG modes present in this pedestal in more detail. \fig{fig:nz0_scan} shows that high parallel resolutions of $n_z=360$ are required to reach fully converged ETG growth rates at high $k_y\rho_i>40$. The parallel structure of the modes can be investigated in the ballooning representation shown in \fig{fig:ballooning}. At $k_y\rho_i=0.2$ (left) and $k_y\rho_i=10, \kxcenter=0$ (middle) the modes have a complex parallel structure, indicative of slab-ETG modes. At a finite ballooning angle $\kxcenter=-35$, however, a toroidal ETG mode with a simpler parallel structure is present for the same binormal wavenumber $k_y\rho_i=10$. The $\kxcenter=-35$ mode is the fastest growing mode at $k_y\rho_i=10$. At smaller scales $k_y\rho_i=100$ (right) the ETG modes are strongly ballooned, with a simple parallel structure, typical for toroidal ETG modes. Note that the x-axes in \fig{fig:ballooning} change from panel to panel. The modes become narrower in parallel direction, as the binormal wavenumber increases. These mode structures demonstrate the existence of a complex mixture of ETG modes in the hybrid H-mode pedestal at different binormal wavenumbers as found in other JET pedestals \cite{parisi_toroidal_2020,chapman-oplopoiou_role_2022}. Our investigation furthermore highlights the existence of toroidal and slab-ETG modes at the same binormal wavenumber, but different ballooning angles.

\subsection{Turbulent ion-scale heat flux}
\label{sec:nonlinear}
\begin{figure}
    \centering
    \includegraphics[width=\linewidth]{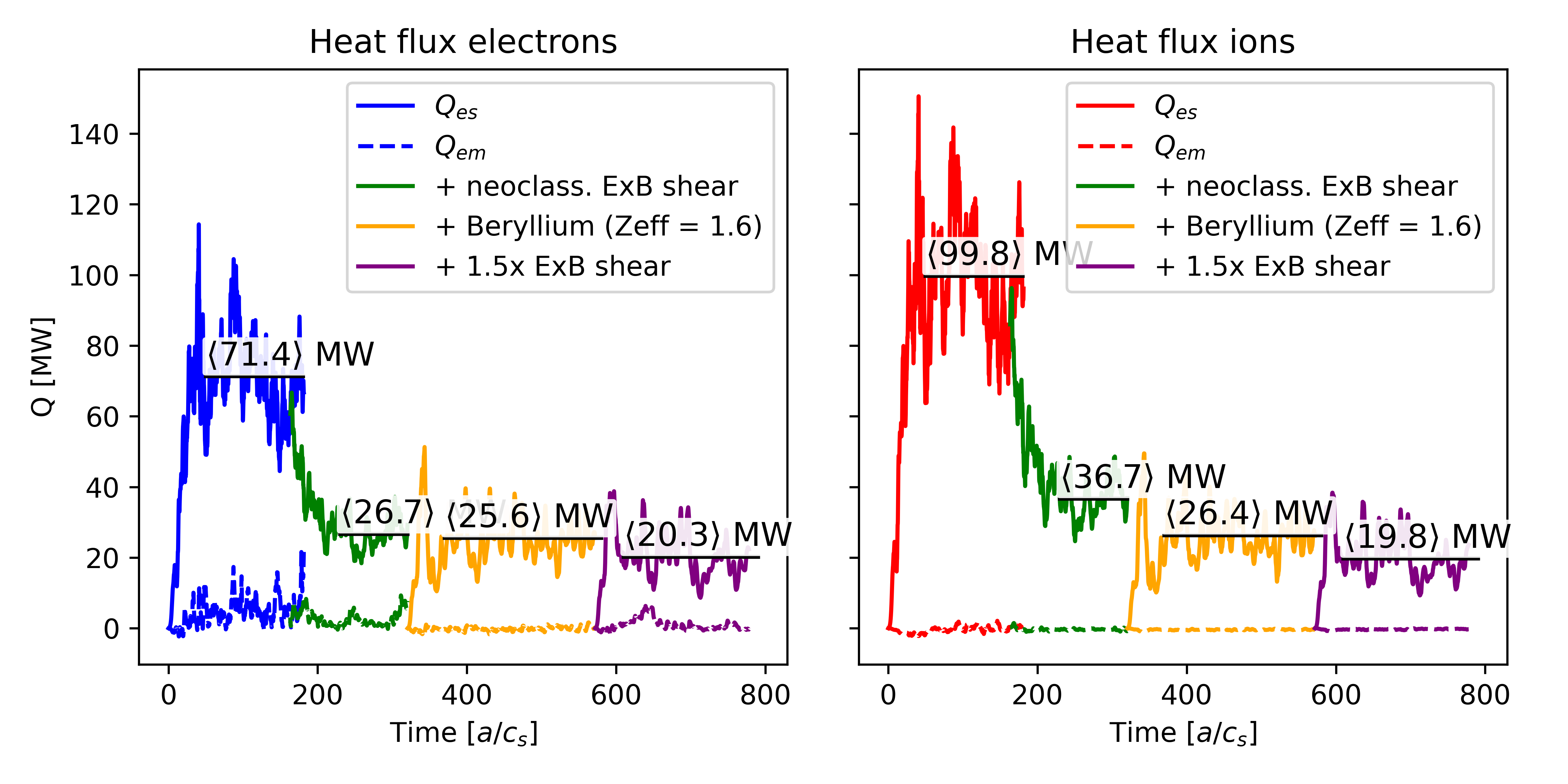}
    \caption{Heat flux as a function of time, averaged over flux surface and radius $\rho_{tor}=0.85 - 0.99$.}
    \label{fig:heatflux_nominal}
\end{figure}

Global, nonlinear simulations reveal the structure of turbulent fluxes in the pedestal. Fig.~\ref{fig:heatflux_nominal} shows the heat flux (averaged over flux surface and radius $\rho_{tor}=0.85 - 0.99$) of electrons (left) and ions (right) as a function of time. The simulations are started with two species (electrons, deuterium) and no $E\times B$ shear (blue and red). Then $E\times B$ shear (cf.~\fig{fig:vrot}) is included (green), which strongly reduces both the electron and ion heat flux channel. Next, a three-species simulation (electrons, Deuterium, Beryllium) tests the influence of an impurity species on the heat flux (yellow). The electron channel is unaffected, but the main ion heat flux is reduced. Finally, the sensitivity to the $E\times B$ shear amplitude is tested by increasing $E\times B$ shear by a factor of 1.5 (purple). Both channels are slightly reduced. This underlines the importance of $E\times B$ shear on the total heat flux level and of impurities on the main-ion-to-electron heat flux ratio. In all simulations, electrostatic heat flux greatly exceeds the electromagnetic component. The main ion channel carries about 60\% of the heat flux, reducing to 50\% when impurities are included. It is important to note that the turbulent fluxes show strong radial variations, as discussed in Fig.~\ref{fig:heatflux}, which are not resolved in Fig.~\ref{fig:heatflux_nominal}.

Experimentally the investigated shot had a total heating power of 33 MW and radiation losses of about 12 MW \cite{hobirk_jet_2023}. Hence, 21 MW should be transported by turbulent and neoclassic heat flux channels on average. The turbulent fluxes in the analyzed radial domain exceed the experimentally inferred fluxes by a factor of two, even if $E\times B$ shear and impurities are included in the simulations. The influence of gradient uncertainties is addressed in Section \ref{sec:variation}.

\subsection{Quasi-linear aspects: Cross-phase and frequency comparison}
The heat flux composition (mostly electrostatic, $Q_i>Q_e$) discussed in the previous subsection, is consistent with the linear growth rate spectrum (ITG \& ETG). To inform the development of reduced transport models for the pedestal, it is interesting to examine the relation between fastest growing modes and nonlinear turbulent state in more detail. The cross-phases of the fluctuating fields and the mode frequencies are well-suited quantities for this analysis.

\begin{figure}
    \centering
    $\rhotor=0.89$
    \includegraphics[width=0.9\linewidth]{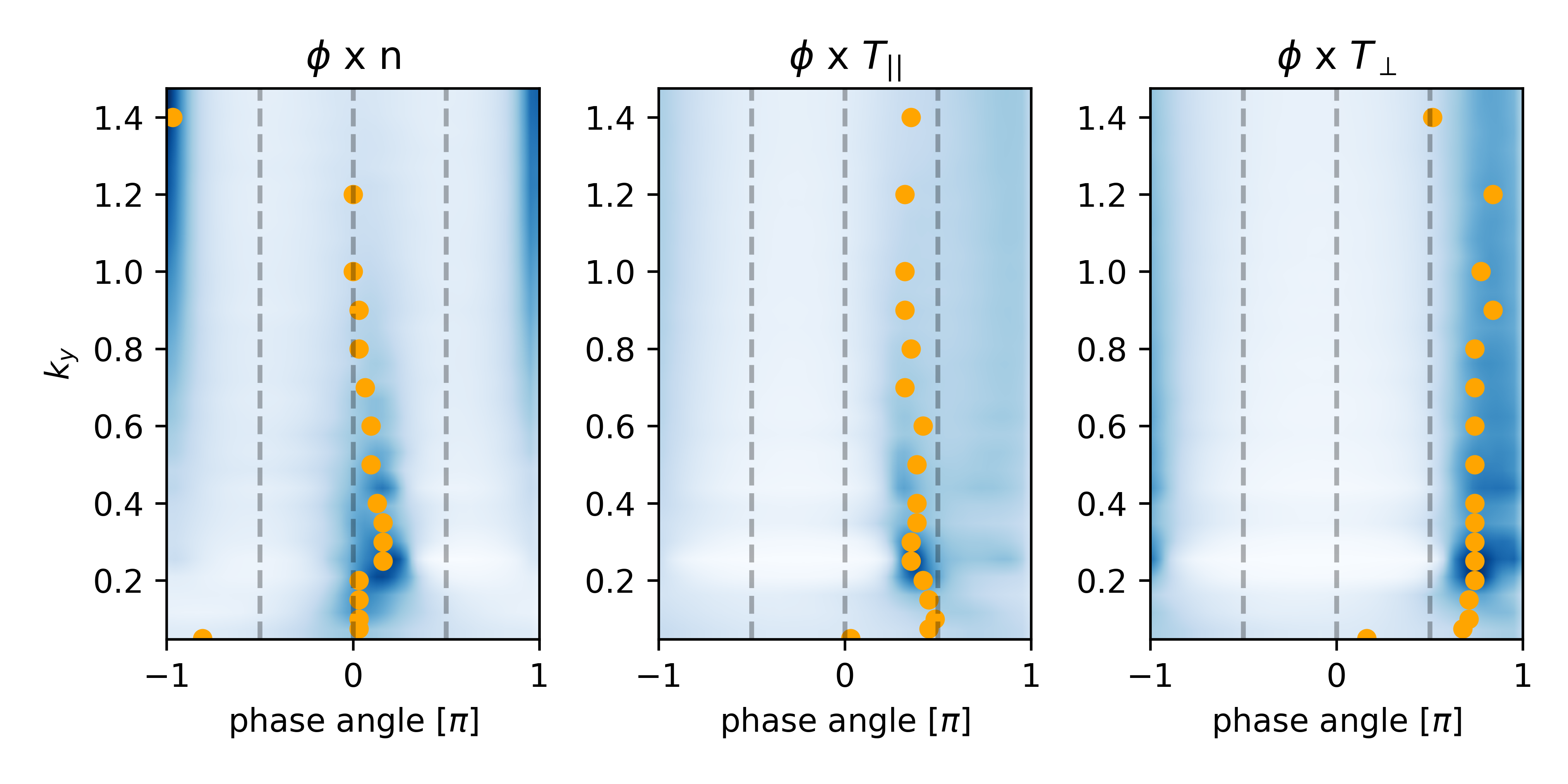}
    $\rhotor=0.97$
    \includegraphics[width=0.9\linewidth]{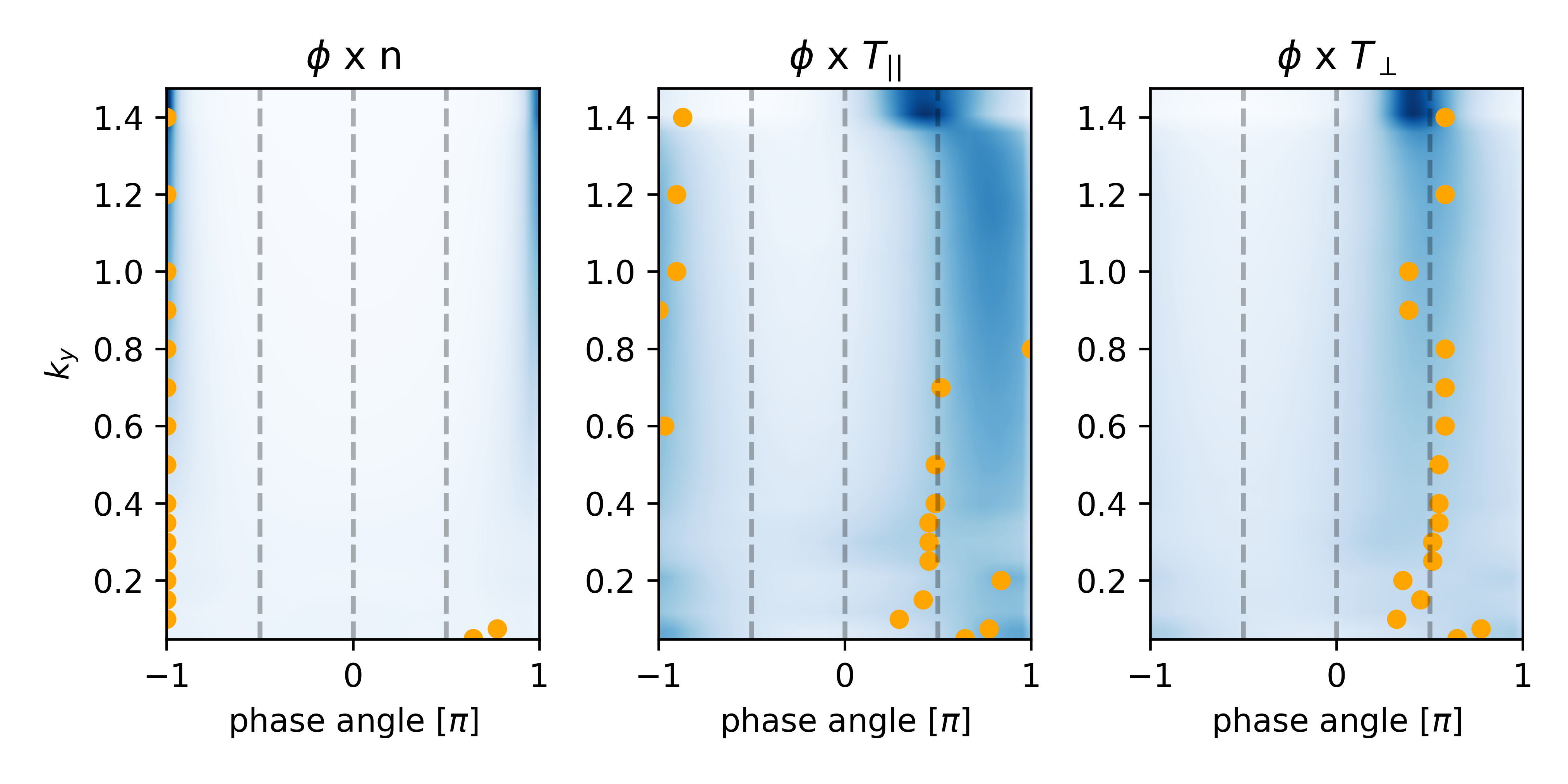}
    \caption{Cross-phases of density and parallel and perpendicular temperature with electric potential at the pedestal top $\rhotor=0.89$ (top row) and pedestal center $\rhotor=0.97$ (bottom row). 
    The blue background shows the nonlinear distribution and linear results are shown as orange circles. The nonlinear simulation with three species and $E\times B$ shear was analyzed (see orange line in \fig{fig:heatflux_nominal}) in the time interval 50-194, corresponding to the black average on top of the orange line.}
    \label{fig:crossphase}
\end{figure}
\fig{fig:crossphase} compares cross-phases of modes from the linear simulations (orange circles) with the distribution in the nonlinear simulations (blue background) analyzed at the pedestal top (top row) and pedestal center (bottom row). In particular, at the pedestal top, a good agreement can be observed. But also at the pedestal center $\phi \times T_\perp$ seems to retain linear structures in the nonlinear state. The cross-phases of $\phi \times n$ give insight into the particle transport. At the pedestal top, the dominant mode has a finite $\phi \times n$ at $k_y\rho_i=0.3$, which indicates transport of particles (transport is maximal at a phase difference of $\pi/2$). This corroborates the dominance of ITG transport at the pedestal top derived from the linear analysis since ITG modes typically cause particle transport.  
In the pedestal center, however, the $\phi \times n$ cross-phase indicates vanishing particle transport. This is in agreement with the transport driver being ETG in the pedestal center.

\begin{figure}
    \centering
    \includegraphics[width=\linewidth]{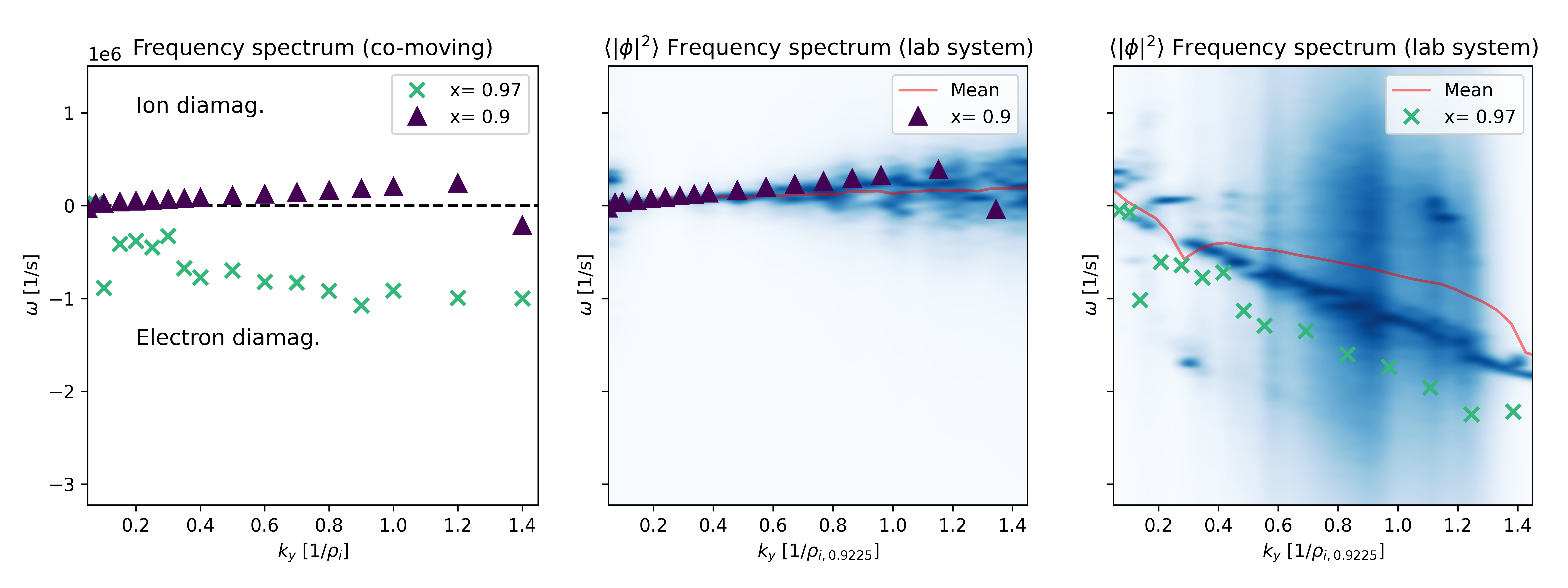}
    \caption{Linear-nonlinear frequency comparison. Left: Linear results in the co-moving frame for two positions. Middle: Comparison of linear (purple triangles) and nonlinear frequencies (blue distribution, analyzed in $\rhotor$=[0.89,0.91]) at pedestal top. Right: Comparison of linear (green crosses) and nonlinear frequencies (blue distribution, analyzed in $\rhotor$=[0.96,0.98]) at pedestal center. The red line indicates the mean of the nonlinear frequency distribution. Analyzed was the nonlinear simulation with three species and $E\times B$ shear (see orange line in \fig{fig:heatflux_nominal}) in the time interval 50-194, corresponding to the black average on top of the orange line.}
    \label{fig:frequency}
\end{figure}
\fig{fig:frequency} compares frequencies from the linear simulations (triangles and crosses) with the distribution in the nonlinear simulations (blue background) analyzed at the pedestal top (middle) and pedestal center (right). The leftmost plot is in the co-moving frame and allows one to distinguish between ion and electron diamagnetic drift directions. The other plots show the nonlinear frequency distribution as a blue background at two selected positions (pedestal top and center) overlaid by frequencies from the linear scan. For the comparison, the linear frequencies are shifted to the lab frame, taking into account the additional $E\times B$ drift due to the radial electric field, which shifts the apparent frequencies. At the pedestal top, linear and nonlinear frequencies coincide very well - the nonlinear state retains linear characteristics. In the pedestal center, the match is worse, but some linear properties seem to survive as well. However, the nonlinear frequency distribution at the pedestal center has a much greater variance than the pedestal top distribution.

These results suggest that the turbulent state at the pedestal top and center retains properties of the linearly fastest growing modes  encouraging the further development of quasi-linear models for turbulent transport in the pedestal.

\subsection{ETG heat flux}
The global, nonlinear simulations of the previous sections only included modes up to $k_y\rho_i\approx1.4$. The linear growth rate analysis, however, shows that there is a broad range of unstable ETG modes at smaller scales. Furthermore, recent pedestal studies show that such ETG modes can drive significant heat flux in the pedestal \cite{chapman-oplopoiou_role_2022, hatch_reduced_2022, leppin_complex_2023-1}. Therefore, we perform local, nonlinear simulations on ETG scales ($k_y\rho_i=[3,192]$) at $\rhotor=0.97$ to assess the small-scale ETG contribution to the total heat flux in the JET hybrid H-mode pedestal. 

The local, linear analysis showed that large parts of the ETG spectrum have slab-ETG mode contributions, which possess a complex parallel structure requiring a high parallel resolution. These fine mode structures reflect in demanding nonlinear ETG simulations. We therefore perform a scan in parallel resolution and hyperdiffusion \cite{pueschel_role_2010} at one radial position (pedestal center, $\rhotor=0.97$). We then compare the obtained heat flux to a recently proposed quasi-linear ETG model \cite{hatch_reduced_2022} to infer the ETG heat flux at other locations in the pedestal.

\begin{figure}
    \centering
    \includegraphics[width=\linewidth]{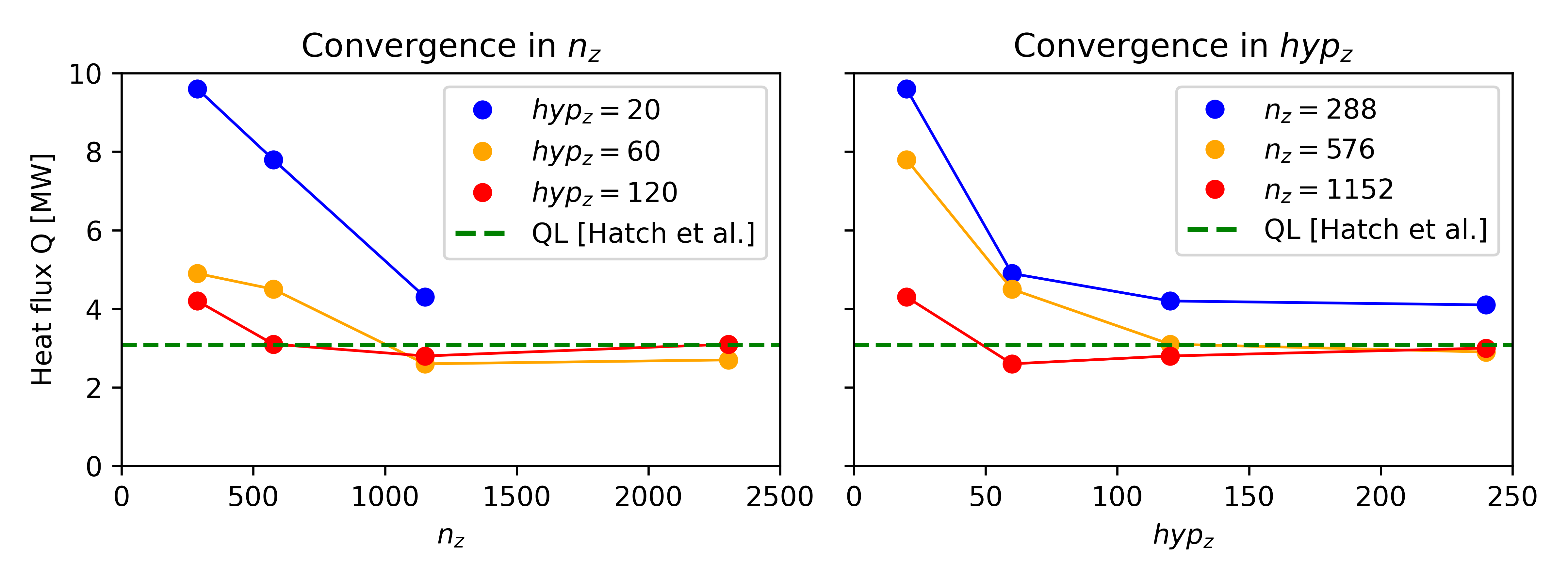}
    \caption{Convergence of ETG heat flux in parallel resolution (left) and parallel hyperdiffusion (right).}
    \label{fig:etg_conv}
\end{figure}

\begin{figure}
    \centering
    \includegraphics[width=0.45\linewidth]{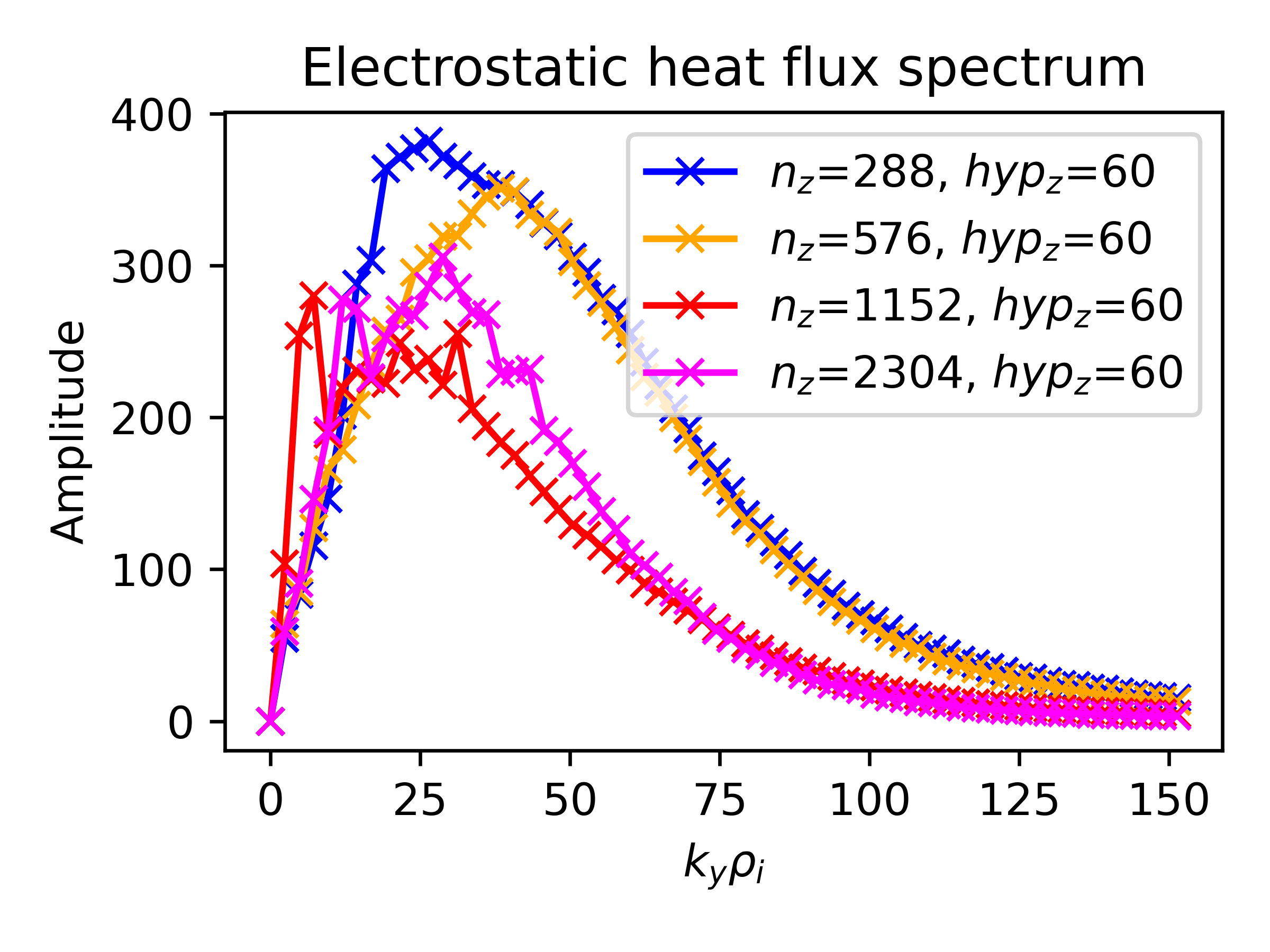}\includegraphics[width=0.45\linewidth]{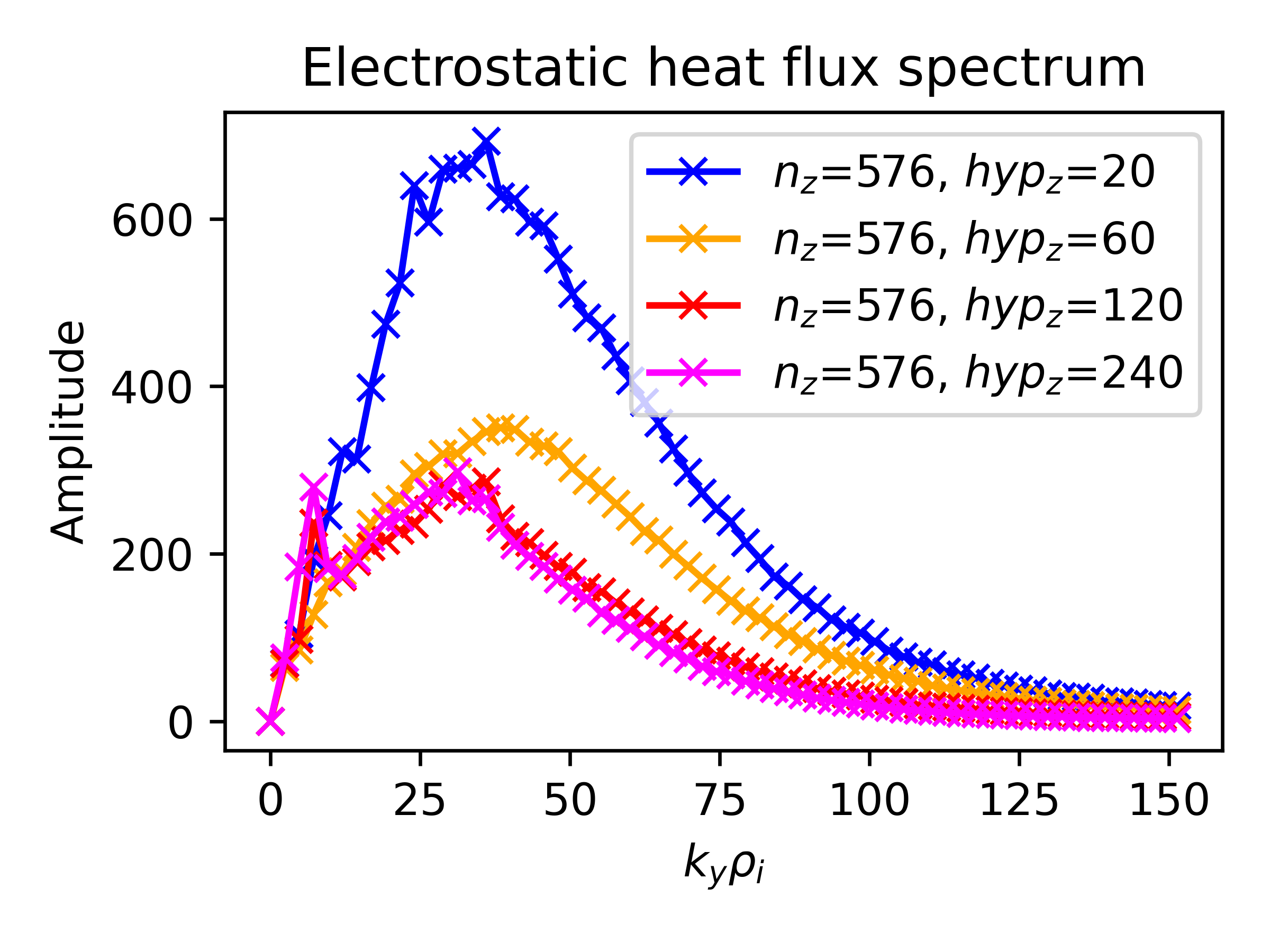}
    \caption{Influence of parallel resolution and hyperdiffusion on the ETG heat flux spectrum. Left: Fixed $hyp_z=60$. Right: Fixed $n_z=576$.}
    \label{fig:etg_spectrum}
\end{figure}

\fig{fig:etg_conv} shows the obtained heat fluxes as a function of the number of parallel grid points $n_z$ (left) and parallel hyperdiffusion strength $hyp_z$ (right). Parallel hyperdiffusion is used in GENE to prevent the aliasing of modes in parallel derivatives and the associated occurrence of grid-scale oscillations \cite{pueschel_role_2010}. Convergence in resolution is only reached with $n_z=1152$ grid points. Convergence in hyperdiffusion strength depends on the resolution, requiring less hyperdiffusion at higher resolution for converged results. The converged heat flux for this shot matches very well with the quasi-linear model by Hatch et al. \cite{hatch_reduced_2022}. We therefore use this model to estimate small-scale ETG contributions to total heat flux at other radial positions. Resolution and hyperdiffusion strength significantly affect the heat flux spectrum as is shown in \fig{fig:etg_spectrum}. The hyperdiffusion strength is specified for a GENE simulation with a mass normalization $\hat{m}_D=1$.

The heat flux decreases with increasing resolution and hyperdiffusion strength before reaching convergence. A visual inspection of the mode contours suggests the following mechanisms behind these trends: Generally, high radial heat flux is associated with wide coherent structures. At too low resolutions the smallest size of these eddies is limited by the grid resolution, resulting in too large eddies, causing too much transport. As the resolution increases, the fine structure of the previously underresolved eddies is revealed, reducing the size of the coherent structures and hence the transport. The parallel hyperdiffusion, on the other hand, smoothes out the eddies in the parallel direction, effectively shortening them radially and hence reducing transport. 

\subsection{Radial heat and particle flux structure}
\begin{figure}
    \centering
    \includegraphics[width=0.75\linewidth]{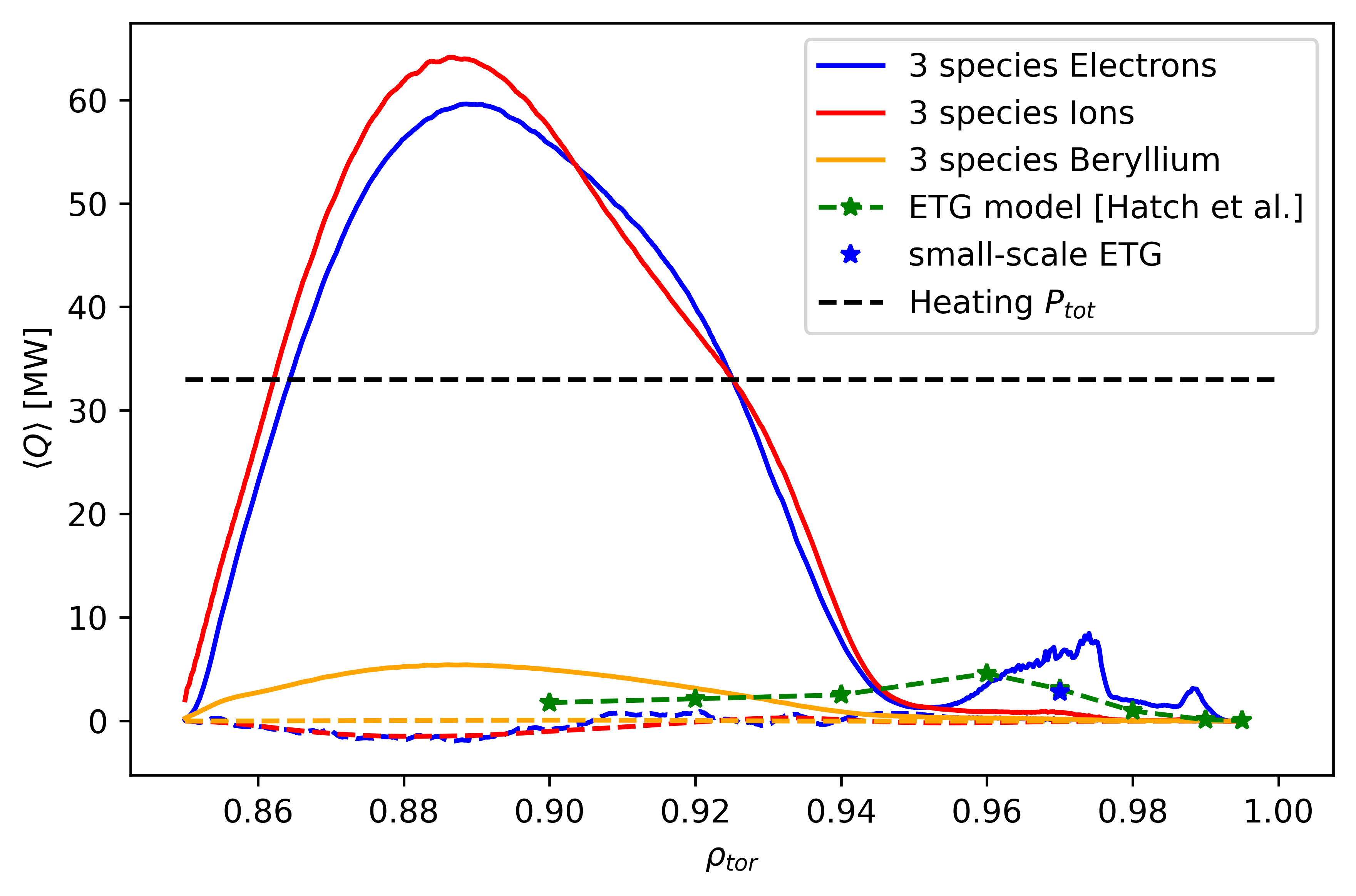}
    \caption{Radial heat flux structure of the global, nonlinear simulation with three species and $E\times B$ shear. On both sides of the simulation domain buffer zones covering 5\% of the domain dampen fluctuations towards the Dirichlet boundary conditions.}
    \label{fig:heatflux}
\end{figure}

The strong changes in the pedestal, require a radially resolved analysis of the turbulent heat flux. \fig{fig:heatflux} shows the heat flux profile for the global nonlinear three species simulation with $E\times B$ shear (cf.~orange time trace in Fig.~\ref{fig:heatflux_nominal}). The simulations with two species and lower $E\times B$ shear show a very similar structure at different heat flux levels. At the pedestal top a strong heat flux peak due ITG modes is visible. It is limited to the left by the inner radial Dirichlet boundary condition of the domain and falls off to the right until it reaches a minimum at $\rhotor=0.95$. On both radial boundaries buffer zones covering 5\% of the domain each dampen fluctuations towards the Dirichlet boundary conditions. Gyroradius effects extend the influence of the buffer zones on both sides, but more strongly on the inner boundary due to the larger gyroradii. In the pedestal center, the turbulent electron heat flux rises again. It has contributions from the low-$k_y$ ETG modes in the global simulation (blue line) and additional contributions from high-$k_y$ ETG modes (green dashed line). Even though this is the simulation with the lowest heat fluxes (strong $E\times B$ shear, inclusion of impurities), the peak heat fluxes that are reached in the pedestal top/ outer core region ($\rhotor\approx0.9$) are unreasonably high with $Q_{tot}\approx100~\rm{MW}$. This is most likely due to a too-steep assumed ion temperature profile in this region. Simulations with modified ion temperature profiles confirm a strong sensitivity of heat flux levels on the gradients. Additionally, even for $T_e$ and $n_e$ a range of alternative profiles and particularly gradients would be consistent with measurements due to the scatter of experimental data (cf.~Fig.~\ref{fig:profiles}) and impact the obtained fluxes.

The structure of the heat flux is also interesting in comparison to a similar analysis performed on turbulent transport in an AUG pedestal \cite{leppin_complex_2023-1}. As in the AUG case, turbulent ion transport is suppressed in the steep gradient region, beginning from around $\rhotor=0.95$, implying that the total ion transport in this region is predominantly neoclassic. In contrast to the AUG pedestal, an electron transport peak due to ion-scale fluctuations is present at $\rhotor=0.97$ in the steepest gradient region. This indicates that in the JET hybrid H-mode pedestal, an effective ion-scale electron transport mechanism in the pedestal center remains.

\begin{figure}
    \centering
    \includegraphics[width=0.75\linewidth]{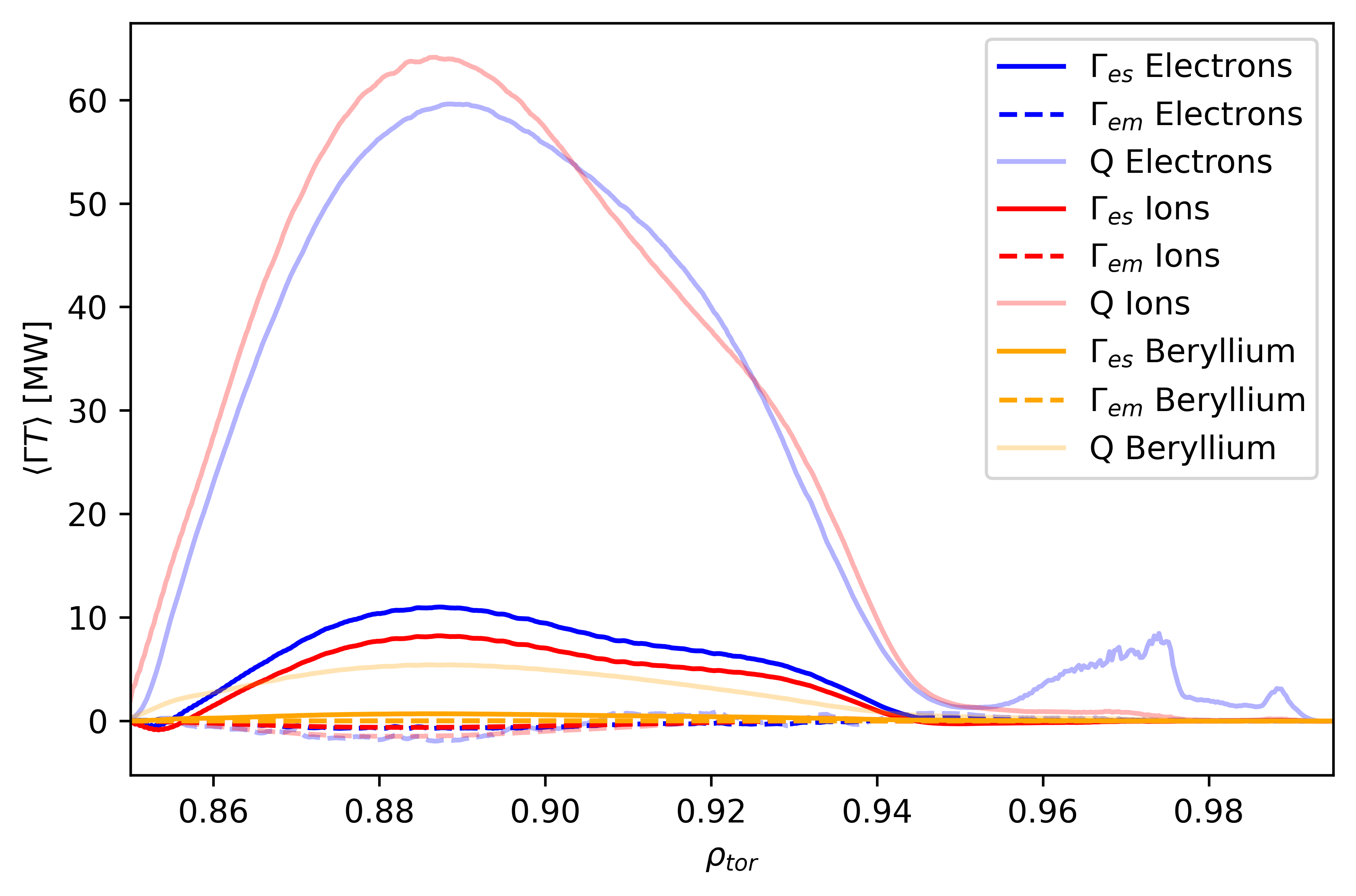}
    \caption{Radial particle flux profile in comparison to heat flux (transparent lines).}
    \label{fig:partflux}
\end{figure}

\fig{fig:partflux} shows the radial structure of the particle flux. It is multiplied by the temperature profile and normalized to MW for a direct comparison with the heat flux level. Significant turbulent particle transport is only present at the pedestal top, being responsible for about 20\% of total turbulent energy loss. In the pedestal center, no turbulent particle transport is observed in the simulations. The radial particle flux structure is consistent with the presence of ITG modes at the pedestal top and ETG modes at the pedestal center.

This concludes the analysis of gyrokinetic instabilities and fluxes at nominal parameters in the pedestal of JET hybrid H-mode \#97781. The next section investigates the impact of global density and temperature profile variations on the pedestal turbulence. The variations are designed to contribute to questions related to the improved confinement properties of hybrid H-modes and their viability for ITER.

\section{Investigation of profile variations}
\label{sec:variation}
\subsection{Influence of density profile}
Hybrid discharges at JET are achieved with high heating powers at low densities compared to the baseline scenario. By artificially increasing the density in our scenario to reach values typical for baseline scenarios we can assess if pedestal turbulence is significantly altered by the lowered density of the JET hybrid scenario. By decreasing the density and collisionality further
we test if the observed turbulence characteristics may be extrapolated to ITER conditions or if significant changes are to be expected. 
\begin{figure}
    \centering
    \includegraphics[width=0.6\linewidth]{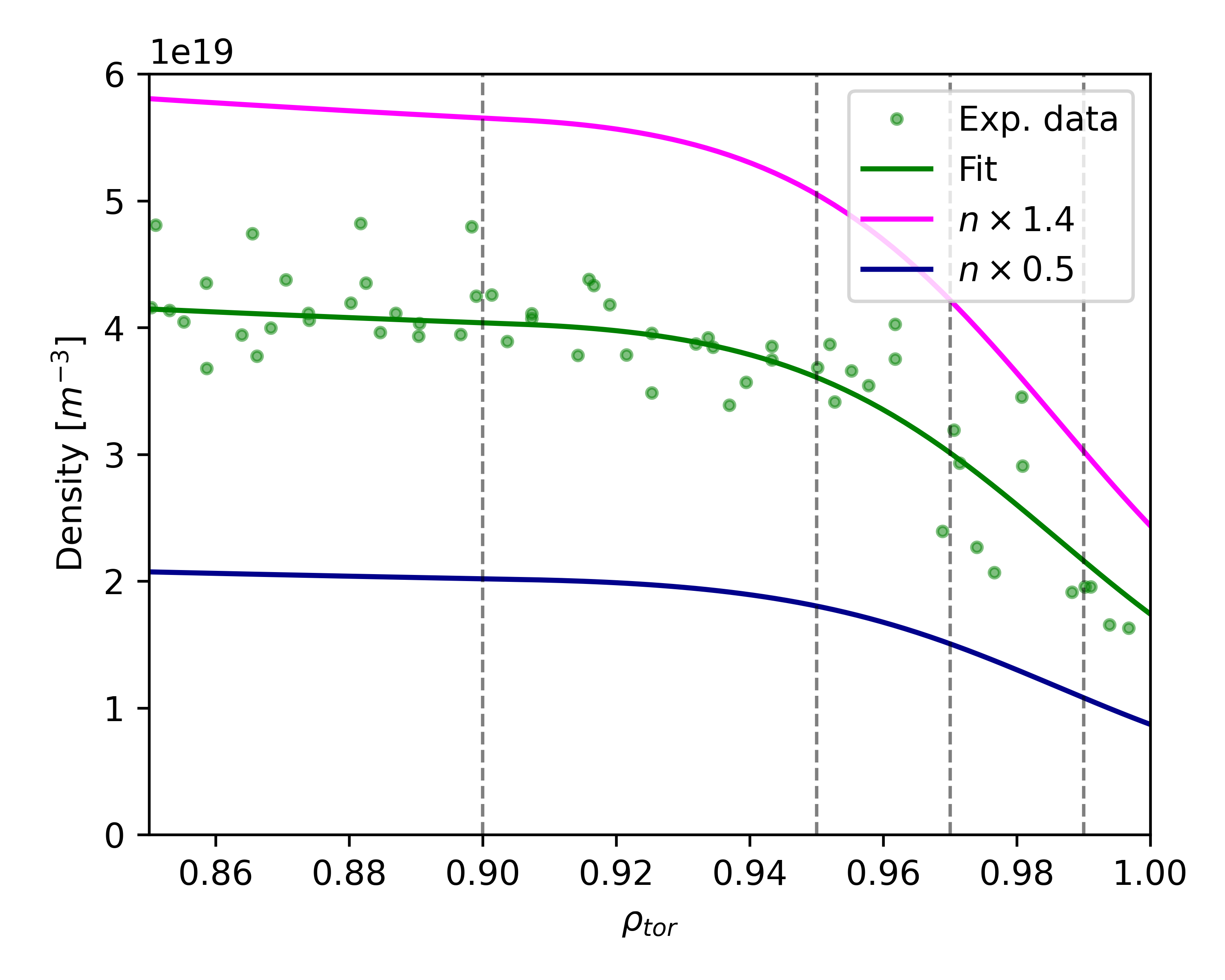}
    \caption{Investigated density profile variations.}
    \label{fig:densvar}
\end{figure}

The profiles are changed such that the turbulent gradient drive is kept constant to study the effect of increased absolute density and collisionality in isolation. This is achieved by multiplying the profiles with a constant $c$. In contrast to e.g. shifting the profiles by adding a constant, this preserves the driving gradient scale length. While the gradient itself is changed by the transformation $\nabla (cn(\rho))=c\nabla n(\rho)$ the gradient length is preserved: $1/L_{cn}=\nabla (cn(\rho))/cn(\rho)=\nabla n(\rho) / n(\rho) = 1/L_n$. The magnetic equilibrium is not altered self-consistently but remains unchanged between profile variations. Hence, the profile changes induce a change in the plasma $\beta$ which is addressed in the following discussion.

\begin{figure}
    \centering
    \includegraphics[width=\linewidth]{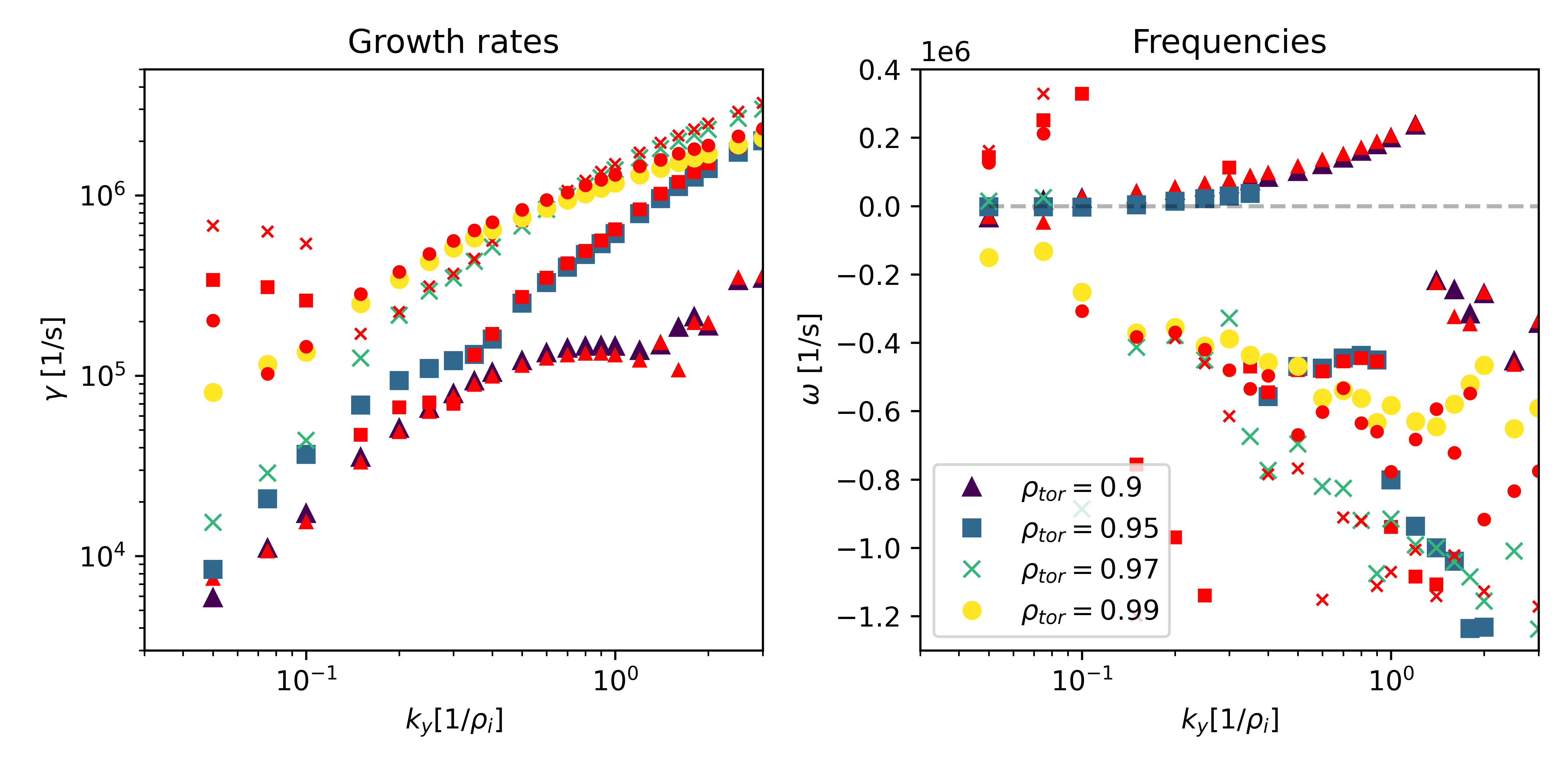}
    \caption{Change in linear growth rates and frequencies with increased density $n\times 1.4$ (red markers) in comparison to results at nominal parameters (non-red markers).}
    \label{fig:linear_nx1p4}
\end{figure}

The impact of the increased density on the linear growth rate spectrum is shown in \fig{fig:linear_nx1p4}. At the positions inwards of the pedestal top $\rhotor=0.95$(squares) and at the pedestal center $\rhotor=0.97$ (crosses) large-scale electromagnetic modes are destabilized. They are identified to be KBMs by their ballooning parity and ion diamagnetic drift direction.

\begin{figure}
    \centering
    \includegraphics[width=\linewidth]{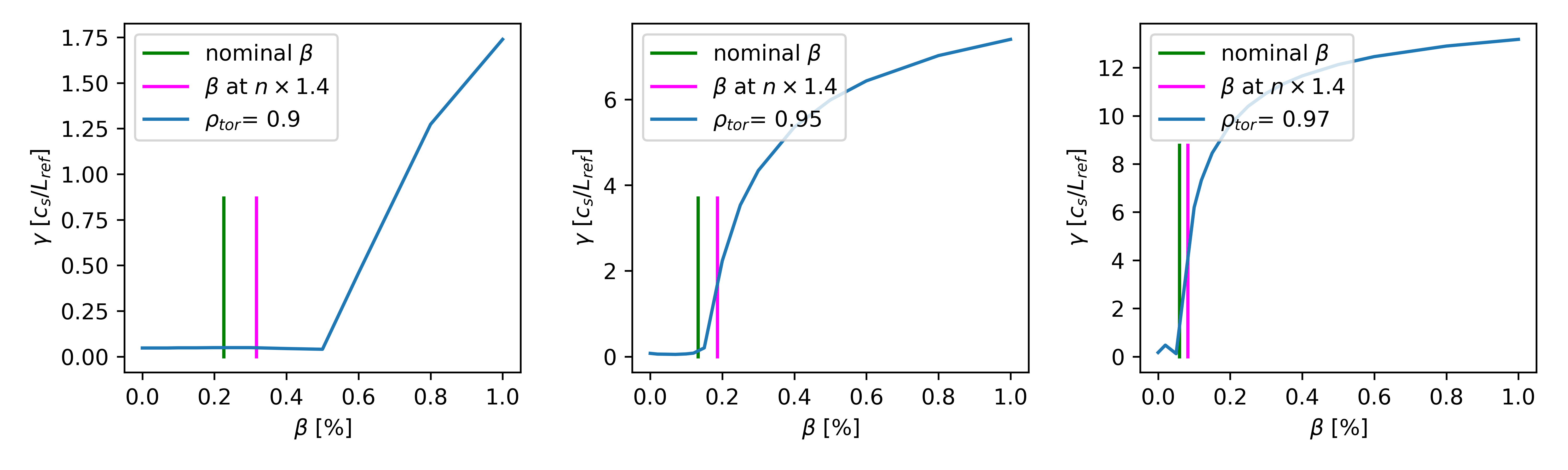}
    \caption{Linear/local plasma $\beta$ scan at three radial positions from pedestal top (left) to center (right).}
    \label{fig:betascan}
\end{figure}

\begin{figure}
    \centering
    \includegraphics[width=\linewidth]{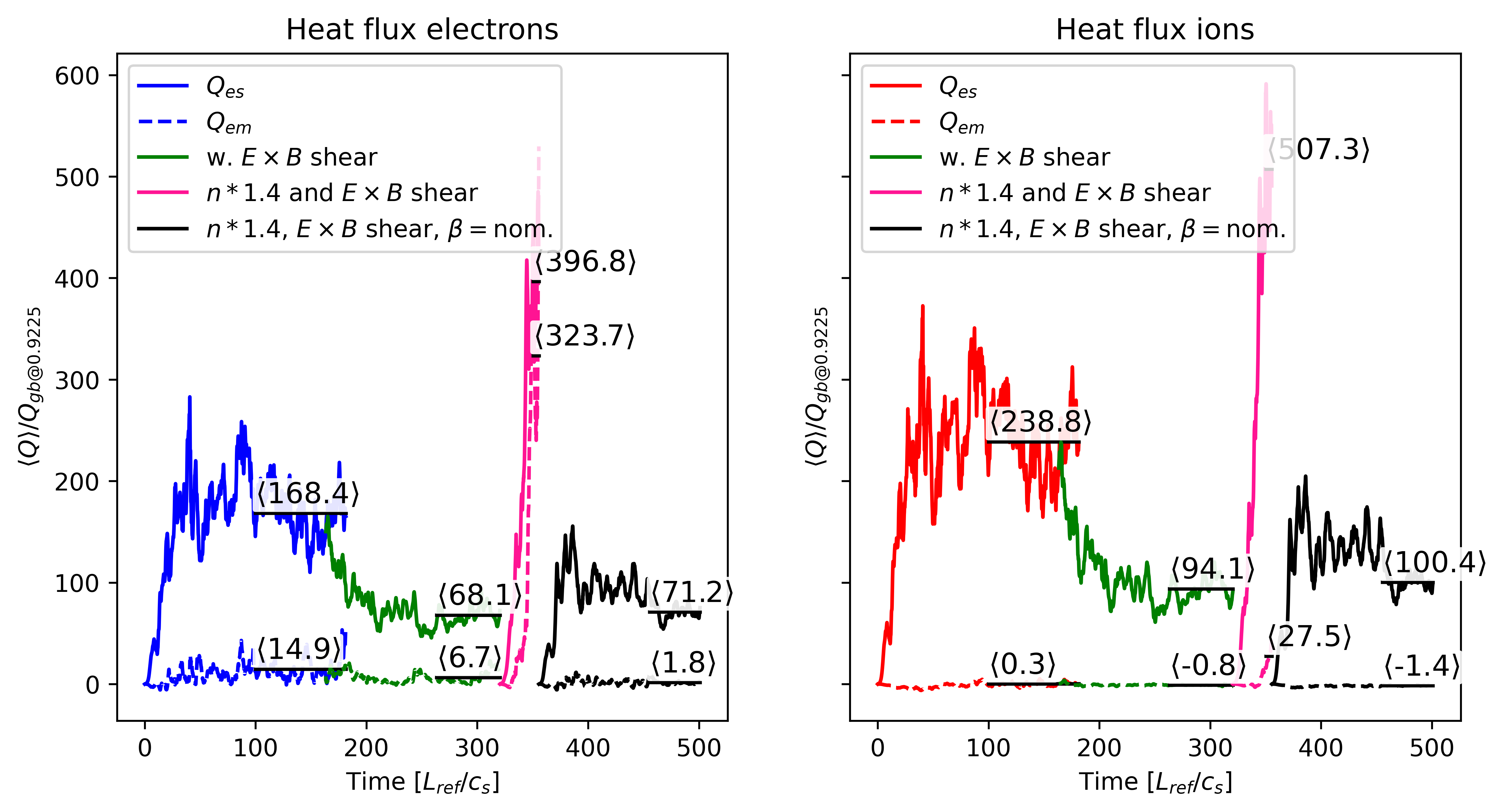}
    \caption{Changes in nonlinear heat flux due to increased density $n\times 1.4$. Including the plasma $\beta$ change (magenta) and keeping $\beta$ artificially constant (black).}
    \label{fig:nonlinear_nx1p4}
\end{figure}

A linear growth rate scan in plasma $\beta$, see \fig{fig:betascan}, confirms that at $\rhotor=0.95$ (middle plot) and $\rhotor=0.97$ (right plot) the density variation crosses a KBM threshold, whereas, at the pedestal top (left plot), the threshold is not crossed. This suggests that particularly around the pedestal center the profiles are limited by electromagnetic KBM modes. In previous analyses of JET pedestals a larger distance to the KBM threshold in the the steep gradient region was found in global/linear simulations \cite{hatch_direct_2019} and in local/linear simulations due to a high bootstrap current \cite{saarelma_mhd_2013}. However, for a $\beta$ increase of x1.4 in the increased density case, global/nonlinear simulations confirm the presence of strong electromagnetic modes, causing the heat flux to diverge (see magenta line in \fig{fig:nonlinear_nx1p4}).

The previous discussion has shown that the turbulent system reacts sensitively to the plasma $\beta$ change induced by the profile variations. It is, however, interesting to investigate if also e.g. the induced change in collisionality would impact the system since a real experimental discharge would operate below the KBM threshold. The increase in plasma pressure by the increased density would be compensated by an increased magnetic field strength. We can model this scenario, by fixing the plasma $\beta$ parameter artificially in the simulations, effectively disentangling the pressure increase from other effects on the turbulent system. We find very little change in the turbulent heat flux when the pressure effect is neglected (compare green and black lines in \fig{fig:nonlinear_nx1p4}). Note that this only holds true for turbulent transport measured in gyro-Bohm units. The transport in SI units changes according to the changing gyro-Bohm normalization, which depends linearly on density $n$.

Overall, the analysis indicates that the reduced pedestal density in the JET hybrid H-mode pedestal is not a decisive factor in improving the confinement of the hybrid H-mode compared to the baseline H-mode. It demonstrates, however, that the pedestal around the center operates close to the onset of a KBM threshold. This suggests the interpretation that the lowered density in the hybrid H-mode pedestal enables the increased ion temperature pedestal, by partially compensating the pressure increase. 

\begin{figure}
    \centering
    \includegraphics[width=\linewidth]{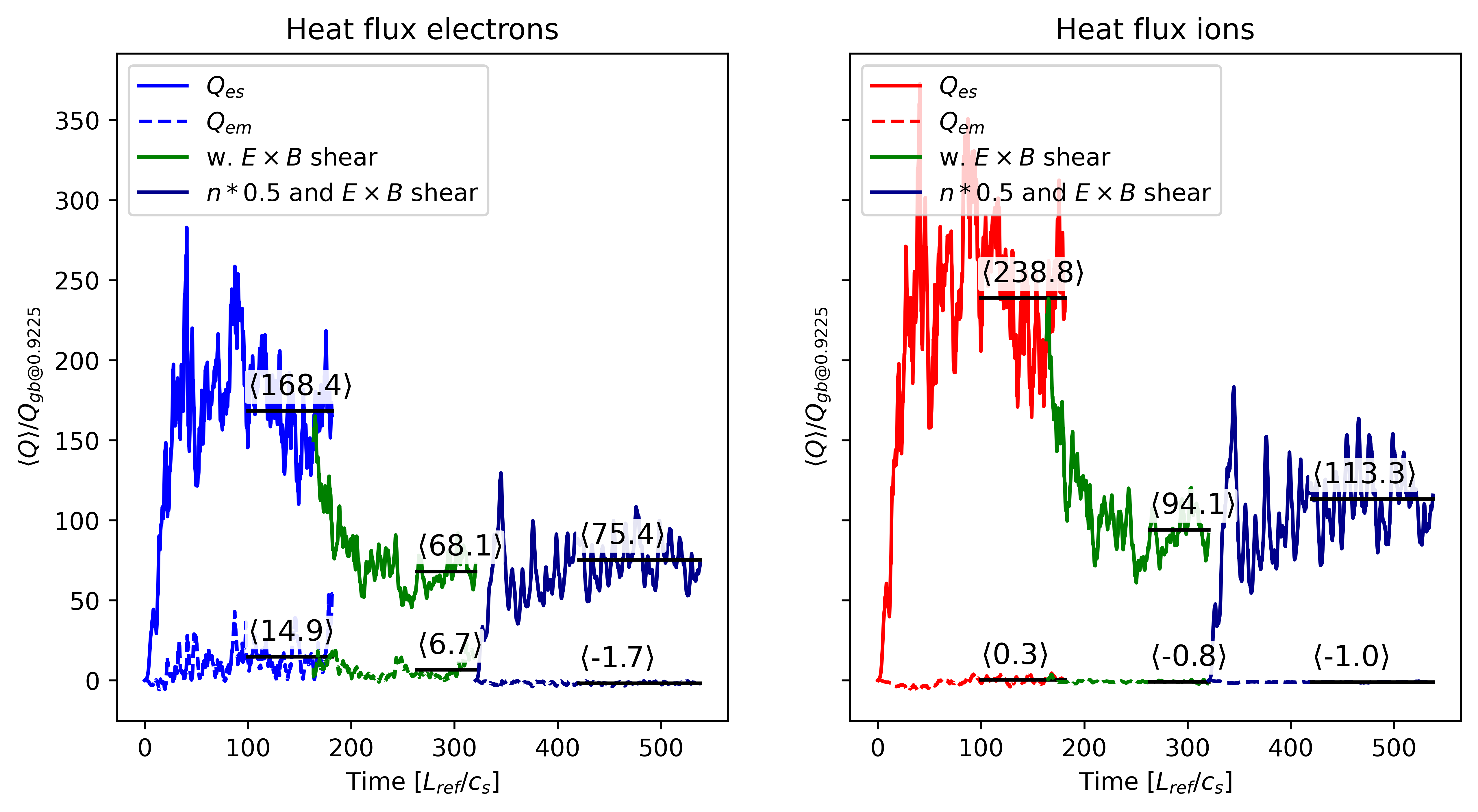}
    \caption{Change in nonlinear heat flux due to lowered density $n\times 0.5$.}
    \label{fig:nonlinear_nx0p5}
\end{figure}

Besides the case with increased density $n\times 1.4$, we also considered the impact of a decreased density $n\times 0.5$ and hence decreased collisionality. \fig{fig:nonlinear_nx0p5} shows the heat flux obtained in nonlinear, global simulations with decreased density. Simulations show a slight increase ($<20\%$) of turbulent heat flux in gyro-Bohm units (compare green and dark blue lines). This suggests that in hybrid H-modes at lower pedestal collisionality no significant, qualitative change in turbulent transport is expected.

\subsection{Influence of ion temperature profile}
\label{sec:temp_variation}
\begin{figure}
    \centering
    \includegraphics[width=0.6\linewidth]{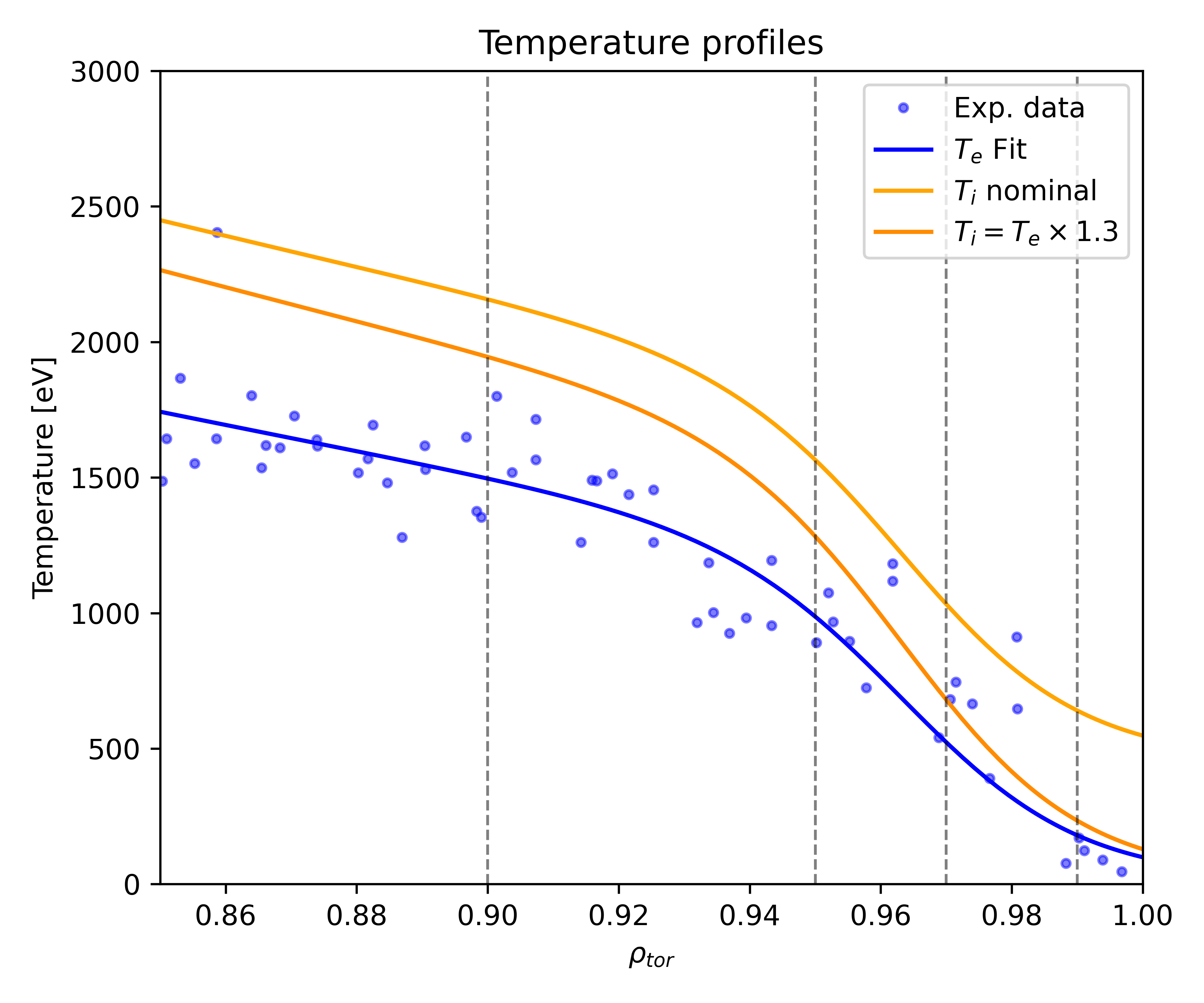}
    \caption{Investigated ion temperature profile variations. $T_i=T_e$ (blue) and $T_i=1.3T_e$ (dark orange). For comparison the nominal estimate for $T_i$ in orange.}
    \label{fig:tempvar}
\end{figure}

Due to the heating setup at JET which includes explicit ion heating by NBI, scenarios can be reached that have rather strongly decoupled ion and electron temperature profiles with $T_i > T_e$. This is the case for the investigated hybrid scenario. In future reactors, however, heating (external and $\alpha$-heating) will pre-dominantly affect the electrons and exhibit more tightly coupled temperature profiles of electrons and ions. This in turn changes the temperature ratio $T_e/T_i$. By comparing the cases of ion temperature $T_i$ equal to electron temperature $T_e$ and $T_i=T_e\times 1.3$ (cf. \fig{fig:tempvar}) we can test if the distinct species temperature profiles are a crucial ingredient of current hybrid discharges from a pedestal turbulence perspective. If they were, this would raise concerns about the viability of the hybrid scenario for future reactors. As explained in the previous section, the change of the profile by a multiplicative constant preserves the gradient scale length, which sets the turbulent gradient drive. The chosen profile variations therefore allow one to study the effect of $T_i>T_e$, without influencing the gradient drive.

\begin{figure}
    \centering
    \includegraphics[width=\linewidth]{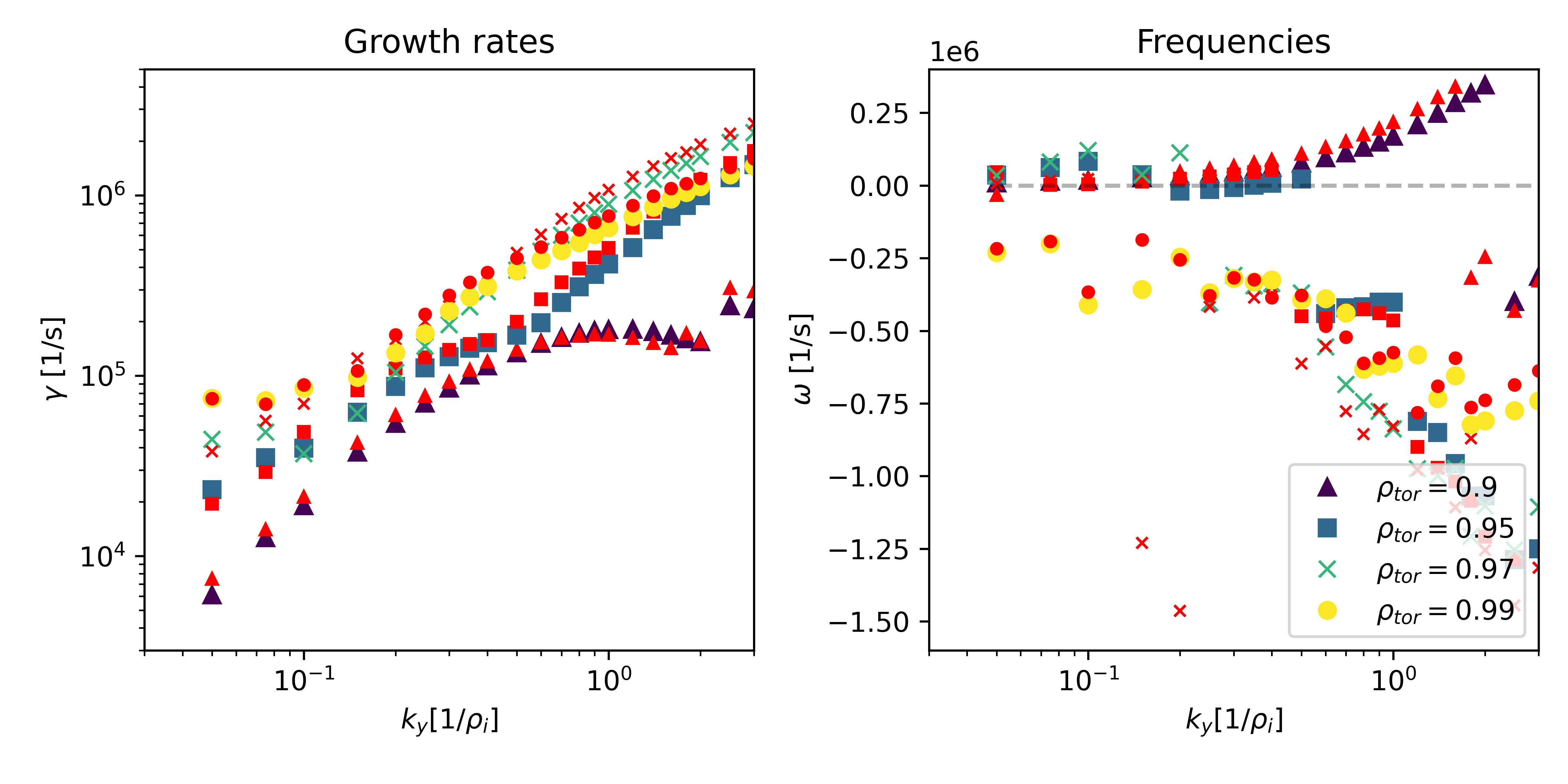}
    \caption{Comparison of linear growth rates and frequencies with $T_i=T_e$ (non-red markers) and $T_i=1.3T_e$ (red markers). }
    \label{fig:linear_temp}
\end{figure}

A comparison of the linear growth rate spectra between $T_i=T_e$ (non-red markers) and $T_i=1.3T_e$ (red markers) is shown in \fig{fig:linear_temp}. At the pedestal top (triangles) growth rates are slightly increased in the $T_i=1.3T_e$ case. Overall, both cases exhibit a qualitatively similar instability spectrum of ITG and ETG modes compared to the spectrum at nominal parameters. A notable exception is the pedestal center $\rhotor=0.97$ on ion-scales (crosses), where at $k_y\rho_i=0.15, 0.2$ an MTM becomes the fastest growing mode in the $T_i=1.3T_e$ case.

\begin{figure}
    \centering
    \includegraphics[width=\linewidth]{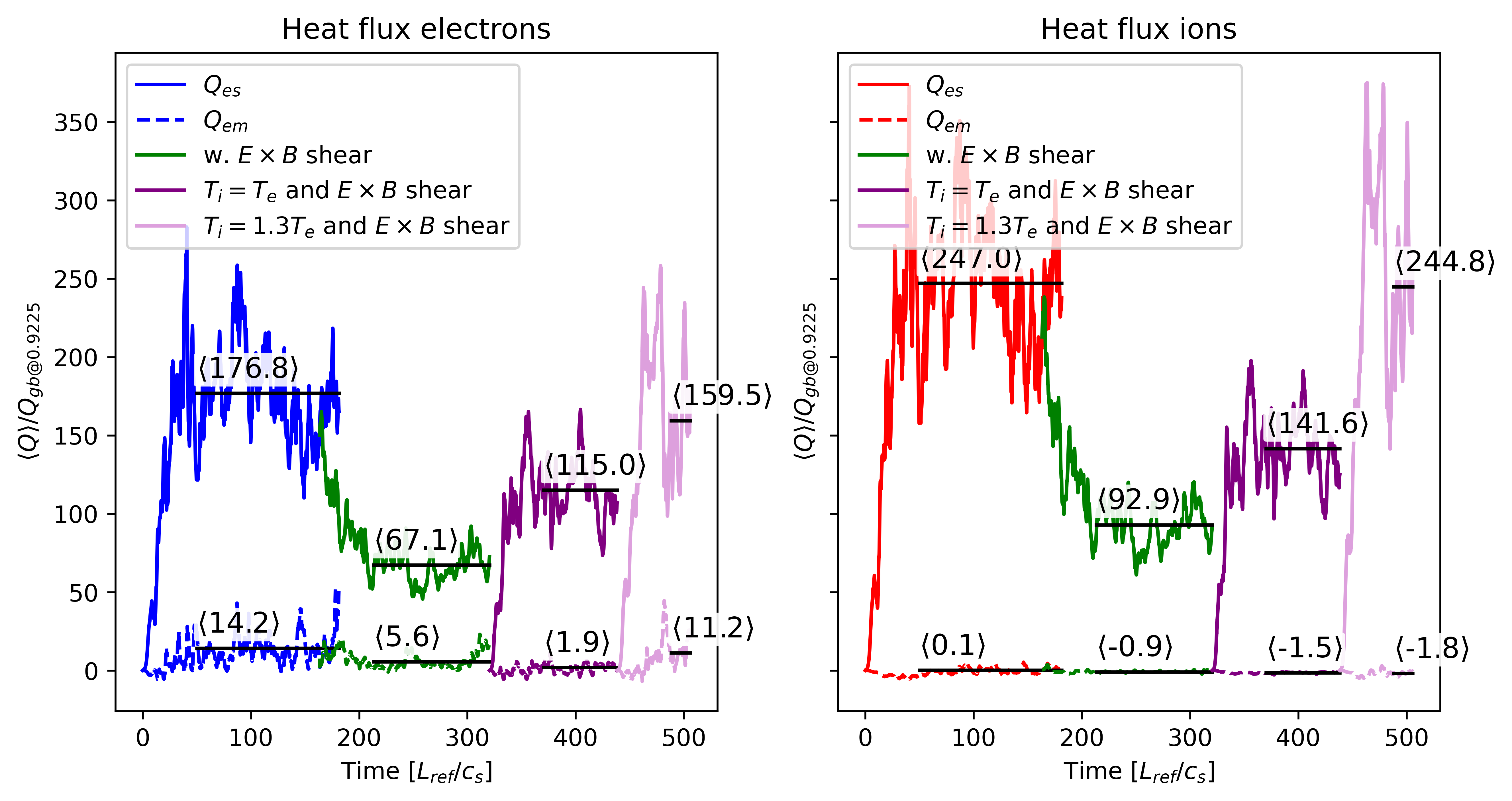}
    \caption{Heat flux comparison between nominal ion temperature profile (green), $T_i=T_e$ (purple) and $T_i=1.3T_e$ (light purple) case.}
    \label{fig:nonlinear_temp}
\end{figure}

\fig{fig:nonlinear_temp} shows the heat fluxes obtained with global, nonlinear simulations for the modified ion temperature profiles $T_i=T_e$ (purple) and $T_i=1.3T_e$ (light purple). The increased ion temperature gradient of the $T_i=T_e$ case in comparison to the nominal $T_i$ profile, reflects in an expected increase in heat flux (compare green and purple lines). The $T_i=1.3T_e$ case shows a further increase in heat flux in the ion and electron channel including a strong relative increase in the electromagnetic, electron channel. This corresponds to the increased pedestal top ITG growth rates and the MTM presence in the pedestal center. 

Based on the comparison of instability spectra and turbulent fluxes obtained with these different ion temperature profiles there are no indications that a scenario with $T_i=T_e$ should have worse turbulent transport properties than a scenario with $T_i>T_e$. In the investigated profile variations, the instability spectrum remained qualitatively unchanged and the $T_i=T_e$ case (reactor-like) produced less turbulent transport than the $T_i=1.3T_e$ case (JET hybrid-like) when comparing cases with the same gradient scale lengths.

\section{Conclusions}
This paper has presented the first comprehensive characterization of gyrokinetic turbulence in a JET hybrid H-mode pedestal. JET discharge \#97781, the deuterium reference discharge in JET's DT hybrid scenario development \cite{hobirk_jet_2023}, was studied. Local/linear, local/nonlinear, and global/nonlinear simulations were performed. The global/nonlinear simulations are among the highest-fidelity pedestal turbulence simulations to date, enabled by a recent upgrade of the global, nonlinear, electromagnetic version of the GENE code \cite{leppin_complex_2023-1}. Dedicated global profile variations were used to explore the sensitivity of the results and inform questions related to the improved confinement properties of hybrid H-modes and their viability for ITER.

Turbulent transport in the investigated pedestal region is mostly driven by ITG and ETG instabilities. The dominant contribution at the pedestal top is ion-scale transport driven by ITGs. Towards the pedestal center and foot, ETG modes become more prevalent, extending down to ion scales. The observed ETG modes have a complex parallel structure with mostly slab-ETG character, resulting in high parallel resolution demands. $E\times B$ shear has been shown to be a crucial component in setting the turbulent heat flux level. Additionally, impurities have been shown to lower the main ion heat transport. Turbulent particle flux is mainly observed at the pedestal top contributing about 20\% to total energy loss. Particularly, the ion temperature profile at the pedestal top is subject to large experimental uncertainties resulting in simulated heat fluxes at nominal parameters larger than the total heating power input.

Quasi-linear properties of the turbulent heat flux were identified in both global, nonlinear simulations and nonlinear ETG simulations. In the global, nonlinear simulations linear mode frequencies and cross-phases could be identified at the pedestal top and center. In the nonlinear ETG simulations a very good match with a recent quasi-linear model \cite{hatch_reduced_2022} is found.

Simulations with modified density profiles suggest that the reduced pedestal density and collisionality in the JET hybrid H-mode pedestal is not a decisive factor in improving confinement compared to the baseline H-mode. The robustness of the turbulent system with respect to collisionality within the studied parameter regime, also indicates that for hybrid H-modes at even lower collisionality, no strong change in the turbulent transport is to be expected. We find, however, that the pedestal is situated close to the onset of a KBM threshold around the pedestal center, limiting the density pedestal through the achievable plasma $\beta$. 

In simulations with a modified ion temperature profile, we do not find a detrimental influence of $T_i=T_e$ compared to $T_i=1.3T_e$ on the turbulent flux level. This suggests that no strong turbulent heat flux increase is to be expected in ITER due to a different $T_i$/$T_e$ ratio compared to present experiments. 

The presented study did not identify principal show-stoppers in the extrapolation of the properties of turbulent transport in hybrid H-mode pedestals to ITER. Neither the lower collisionality nor the altered electron-to-ion temperature ratio produce - in the investigated test cases - strongly increased fluxes or instabilities. This of course does not exclude the possibility that such show-stoppers exist - but within the analysed parameter variations and validity of the employed model they are not present. 

With respect to the improved confinement properties of hybrid H-modes, no principal benefit of the lower collisionality w.r.t. turbulent transport compared to the baseline scenario has been found.

While the presented simulations are among the highest-fidelity gyrokinetic pedestal simulations to date, several limitations apply to the chosen methodology. The simulations are gradient-driven, implying that the underlying density and temperature profiles do not change on average during the simulation. Results are therefore sensitive to experimental measurement uncertainties in the input profiles. The emphasis has been therefore put on studying the influence of physical effects like $E\times B$ shear and impurities on the heat flux, rather than precisely matching experimental heat fluxes by tuning gradients. While the impact of charged impurities has been investigated, neutrals are not included in the present simulations. Neutrals have been shown to influence tokamak edge turbulence \cite{zholobenko_role_2021}. 
Furthermore, the radial simulation domain is restricted to be strictly inside of the separatrix and the magnetic equilibrium has been kept constant for all profile variations. While both ion- and electron-scale simulations were performed, multi-scale effects, i.e. cross-scale interactions between ion-gyroradius size and electron-gyroradius size eddies, are not captured with the presented simulations. Several multi-scale effects influencing the expected heat flux level have been reported \cite{maeyama_cross-scale_2015,howard_multi-scale_2016,howard_multi-scale_2018}.

The discussed limitations suggest natural next steps for future pedestal turbulence simulations, including flux-driven simulations and simulations extending to the near scrape-off layer. Furthermore, the observed quasi-linear properties of the dominant modes show that if growth rates are accurately captured, important properties of nonlinear fluxes may be inferred from linear simulations, encouraging the development of quasi-linear pedestal transport models.

\label{sec:conclusions}

\ack{
The authors thank Gabriele Merlo and Ben Chapman-Oplopoiou for helpful discussions on ETG modes and JET pedestals. 

Simulations for this work have been performed on the HPC system Raven at the Max Planck Computing and Data Facility (MPCDF) and Marconi at CINECA.

This work has been carried out within the framework of the EUROfusion Consortium,
funded by the European Union via the Euratom Research and Training Programme
(Grant Agreement No 101052200 — EUROfusion). Views and opinions expressed
are however those of the author(s) only and do not necessarily reflect those of the
European Union or the European Commission. Neither the European Union nor the
European Commission can be held responsible for them.
}

\section*{Declaration of interest}
Competing interests: The authors declare none. 

\section*{Data availability}
GENE is an open-source code that is freely available via genecode.org upon request. Data and parameter files are available from the corresponding author upon reasonable request.

\section*{References}
\bibliographystyle{iopart-num}
\providecommand{\newblock}{}

\end{document}